\documentclass[preprint,3p,12pt]{elsarticle} 
\biboptions{sort&compress}

\usepackage{amsmath} 
\usepackage{amssymb}
\usepackage{mathrsfs}
\usepackage{graphicx} 
\usepackage{hyperref} 
\usepackage{booktabs}
\usepackage{colortbl}
\usepackage{caption}
\usepackage{subcaption}
\usepackage{multirow}

\begin{document}

\begin{frontmatter}

\title{TransportBench: A Comprehensive Benchmark for Non-Equilibrium Flow Transport} 

\author[imech,ucas1]{Xu Wang\fnref{equal}}
\author[imech]{Minghao Li\fnref{equal}}
\author[imech]{Qizhen Hong}
\author[imech]{Yang Liu}
\author[imech]{Chen-an Zhang}
\author[imech]{Shuai Zhang}
\author[imech,ucas]{Wenhao Li}
\author[imech,ucas]{Yonghao Zhang}
\author[imech,ucas]{Tianbai Xiao\corref{cor1}}
\ead{txiao@imech.ac.cn}

\fntext[equal]{These authors contributed equally to this work.}
\cortext[cor1]{Corresponding author}

\address[imech]{Institute of Mechanics, Chinese Academy of Sciences, Beijing, China}
\address[ucas1]{School of Future Technology, University of Chinese Academy of Sciences, Beijing, China}
\address[ucas]{School of Engineering Science, University of Chinese Academy of Sciences, Beijing, China}

\begin{abstract}
\indent Scientific machine learning models, as versatile tools for numerical simulation and analysis, are increasingly transforming the landscape of fluid mechanics research.
However, existing datasets and benchmarks are primarily limited to continuum fluids and provide limited support for non-equilibrium transport phenomena.
To address this gap, we present TransportBench, a high-fidelity dataset and standardized benchmark for non-equilibrium flow transport, designed to reveal the strengths and limitations of neural network models across diverse flow regimes.

Specifically, the dataset encompasses a broad physical spectrum, covering continuum and rarefied regimes, low-speed and hypersonic flows, inert and chemically reactive gases, and both translational and internal-energy non-equilibrium effects. 
Built upon this dataset, we systematically benchmark representative neural architectures using unified evaluation protocols to probe key challenges in learning non-equilibrium flows, including robustness to shock-dominated discontinuities and multi-scale effects, as well as generalization across geometry and physical parameters.

Numerical results demonstrate that model performance exhibits a pronounced dependence upon the specific flow characteristics.
No single architecture consistently performs best for all the tasks.
Instead, different architectural inductive biases provide distinct advantages in capturing smooth flow fields, shock-induced discontinuities, and high-order non-equilibrium statistics. 
By jointly providing the non-equilibrium flow dataset and model benchmark, TransportBench offers a new testbed for the development, evaluation, and diagnosis of scientific machine learning methods for fluid transport beyond the Navier-Stokes hydrodynamics.
The benchmark datasets\footnote{\url{https://huggingface.co/datasets/CFDML/TransportBench}} and implementation codes\footnote{\url{https://github.com/CFDML/TransportBench}} are available under the MIT license.

\end{abstract}


\end{frontmatter}

\begin{table}
    \centering
    \caption{Nomenclature.}
    \begin{tabular*}{16cm}{lll}
        \hline
        $\mathbf x$ & space variable $(x,y,z)$ \\
        $\mathbf v$ & particle velocity variable $(u,v,w)$ \\
        $t$ & time variable \\
        $f$ & particle distribution function \\
        $s$ & species index \\
        $l$ & internal state index \\
        $\mathcal Q$ & collision operator in kinetic equation \\
        $\rho$ & macroscopic density \\
        $\mathbf u$ & macroscopic velocity \\
        $E$ & macroscopic energy density \\
        $\mathbf U$ & conservative variables of flow field \\
        $\mathbf P$ & pressure tensor \\
        $\mathbf q$ & heat flux \\
        $p$ & macroscopic pressure \\
        $\varphi$ & mass fraction \\
        $h$ & enthalpy \\
        $S$ & source term \\
        $N_s$ &  total number of chemical species \\ 
        $V_s$ & number of vibrational states associated with the s-th species \\ 
        $D_s$ & diffusion coefficient of the s-th species \\ 
        $\eta$ & thermal conductivity associated with translational-rotational energy \\ 
        $\mu$ & shear viscosity \\ 
        $\mathcal A$ & operating conditions \\
        $\mathcal U$ & solution field \\
        $\mathcal G$ & machine learning model \\
        $\mathcal L$ & loss function \\
        $\mathbf b$ & class-shape transformation (CST) coefficient \\
        $\mathscr{U}$ & uniform distribution \\
        $r$ & random variable sampled from $\mathscr{U}$ \\
        Ma & Mach number \\
        Kn & Knudsen number \\
        Re & Reynolds number \\
        $\omega$ & viscosity index\\
        $d_\mathrm{ref}$ & diameter at reference temperature\\
        $\mathcal{T}$ & non-linear logarithmic transformation \\
        $\mathbf M$ & binary geometry mask \\
        $a$ & scalar field \\ 
        $N$ &  number of spatial grid points \\ 
        MSE & mean squared error \\
        MAE & mean absolute error \\
        \hline
    \end{tabular*}
    \label{table:nomenclature}
\end{table}
\newpage



\section{Introduction}

Accurate simulation of non-equilibrium flow transport has become increasingly important as fluid systems are pushed toward more extreme spatial and thermodynamic regimes.
In hypersonic flight and atmospheric re-entry \cite{sziroczak2016review}, gases may experience strong compression, shock heating, rarefaction, and internal-energy excitation; in micro/nano-electro-mechanical systems (MEMS/NEMS) \cite{gad2001mems}, the characteristic length can become comparable to the molecular mean free path.
These conditions give rise to pronounced non-equilibrium effects, under which the continuum assumption may break down, and the classical Navier-Stokes equations become insufficient for accurately describing the underlying transport phenomena.
A more faithful description requires statistical mechanics and gas-kinetic formulations such as the Boltzmann equation \cite{cercignani1988boltzmann}, or direct-modeling particle methods such as the direct simulation Monte Carlo (DSMC) method \cite{bird1994molecular}.
Although these approaches provide reliable descriptions of non-equilibrium fluid dynamics, their high computational cost, slow statistical convergence, and inherent stochastic noise make them difficult to use for rapid design, uncertainty analysis, and repeated parametric studies. This creates a strong demand for efficient surrogate modeling and simulation acceleration methods tailored to non-equilibrium flow transport.


Scientific machine learning is emerging as a promising paradigm for addressing these computational challenges by combining data-driven approximation with physical modeling \cite{brunton2020machine,karniadakis2021physics}.
However, the systematic evaluation of such methods remains largely constrained by benchmarks designed for continuum and near-equilibrium flow problems.
Widely used datasets, such as PDEBench \cite{takamoto2022pdebench} and FlowBench \cite{tali2024flowbench}, primarily consider the Navier-Stokes-level dynamics close to thermodynamic equilibrium, in which the flow fields are often smooth, the output variables are mostly low-order macroscopic quantities, and the computational domains are regular or only mildly complex.
They do not fully reflect the defining challenges of non-equilibrium transport, including rarefaction effects, Knudsen layers, high-order moment quantities, strong shock discontinuities, and multi-scale kinetic-to-continuum behavior.
Consequently, strong performance on existing continuum-flow benchmarks does not necessarily indicate that a model can robustly predict rarefied or hypersonic non-equilibrium flows.


Beyond the issue of physical coverage, existing evaluation methods also leave room for more systematic comparison in non-equilibrium flow learning.
In many studies, neural architectures are assessed under different parameter budgets, grid resolutions, preprocessing strategies, and training protocols. Such differences can make it difficult to distinguish the effect of architectural inductive bias from that of model capacity or implementation details \cite{wang2025fdbench}.
Moreover, different neural architectures encode distinct assumptions about locality, global coupling, mesh structure, and coordinate representation.
These assumptions can strongly affect their ability to resolve sharp gradients, high-order non-equilibrium statistics, and localized flow structures.
In particular, coordinate-based and attention-based models may exhibit spectral bias and tend to learn smoother, low-frequency mappings, while spectral or convolutional models may respond differently to high-frequency feature injection \cite{rahaman2019spectral}.
Therefore, a benchmark for non-equilibrium flow transport should not only report prediction errors, but also provide controlled settings for diagnosing how different architectures behave under specific physical challenges \cite{xiao2021using,xiao2023relaxnet}.


To address these issues, we introduce TransportBench, a benchmark framework with two closely connected objectives. First, it provides a high-fidelity dataset for non-equilibrium flow transport, covering rarefied and continuum regimes, low-speed and hypersonic flows, inert and reacting gases, and both translational and internal-energy non-equilibrium. The dataset is organized into four representative flow scenarios, including airfoil, cylinder, cavity, and double-cone configurations, which are designed to examine geometric generalization, physical-parameter generalization, high-order moment prediction, and shock-dominated hypersonic transport. Second, TransportBench establishes a standardized model benchmark for evaluating representative scientific machine learning architectures under controlled settings. By comparing operator, spectral, convolutional, attention-based, geometric, and latent models with unified preprocessing, training protocols, parameter budgets, and evaluation metrics, the benchmark enables a more systematic assessment of how different architectural inductive biases perform across distinct flow regimes. Through this dual design, TransportBench aims to serve not only as a reusable dataset for non-equilibrium fluid dynamics, but also as a diagnostic testbed for developing and evaluating learning-based methods for flow transport beyond the classical Navier-Stokes level.


\section{Related Work}

\subsection{Neural architectures for learning fluid dynamics}

Scientific machine learning has introduced a broad range of neural architectures for approximating fluid dynamics across different spatial representations and physical scenarios. These models differ not only in the network design, but also in the inductive biases they impose on the learned solution fields. Understanding these biases is essential for evaluating their suitability for non-equilibrium transport problems, where the target fields may contain smooth regions, sharp discontinuities, multi-scale structures, and high-order kinetic statistics.

\textbf{Convolutional and latent-representation models}, such as convolutional neural networks, U-Net \cite{ronneberger2015u}, and autoencoders \cite{zhai2018autoencoder}, have been widely used for learning solution fields represented on structured grids. Convolutional models exploit local receptive fields, hierarchical feature extraction, and translational equivariance, making them effective for capturing spatial gradients and multi-scale coherent structures.
U-Net, in particular, has become a common baseline for image-like flow-field prediction because the encoder-decoder structure and skip connections combine local details with broader contextual information.
Convolutional autoencoders (AE) further introduce a latent bottleneck for compact flow-field representation and are often used for reduced-order modeling and manifold learning.


\textbf{Operator-learning models}, such as the deep operator network (DeepONet) \cite{lu2021learning} and Fourier neural operator (FNO) \cite{li2020fourier}, provide another important route for data-driven fluid simulation. Rather than learning only a finite-dimensional mapping between fixed discretized inputs and outputs, these models aim to approximate mappings between function spaces. 
DeepONet represents nonlinear operators through a branch-trunk architecture and supports continuous spatial queries, while FNO parameterizes integral kernels in the Fourier domain and efficiently captures long-range correlations on structured grids. 
These properties make operator-learning models attractive for learning input-output mappings between operating conditions, geometries, and flow fields, regardless of whether the underlying data are generated by parametric PDE solvers, particle simulations, or experiments.
Their performance, however, may depend strongly on whether the target flow structures are smooth, globally correlated, or dominated by localized discontinuities.

\textbf{Attention-based and geometric learning models} provide flexible ways to model spatial interactions beyond purely local convolutional kernels. Vision Transformers (ViT) \cite{han2022survey} use patch embeddings and self-attention to capture long-range dependencies on structured grids, making them suitable for flow fields where global interactions are important. In contrast, point-based and graph-based models are designed for irregular spatial representations, such as point clouds or unstructured meshes. Point Transformer (PT) and related geometric architectures can directly aggregate information from neighboring points, thereby providing a natural representation of complex geometries without relying entirely on Cartesian grid interpolation \cite{zhao2021point}. Although these models offer greater flexibility in spatial representation and interaction modeling, their effectiveness in sparse, shock-dominated, or highly non-equilibrium regimes remains problem-dependent.

Overall, existing neural architectures provide complementary strengths for learning fluid dynamics. Convolutional models rely on local receptive fields and hierarchical feature extraction to capture spatial gradients and multi-scale structures. Operator-learning models are designed to learn mappings between functions or fields, with Fourier-based variants further exploiting spectral representations to capture long-range correlations. Attention-based models use token-level interactions to model global dependencies, while geometric models operate on point clouds, graphs, or meshes to accommodate flexible spatial representations.
These strengths, however, can also become limitations when the target flow fields contain strong discontinuities or high-order non-equilibrium structures. Coordinate-based models, and in some cases attention-based models, may favor smoother or lower-frequency representations, whereas Fourier-based spectral representations may be prone to oscillatory artifacts near sharp discontinuities. Therefore, a systematic comparison of these architectures under non-equilibrium flow conditions is necessary for understanding which model families are suitable for different regimes.


\subsection{Benchmarks for scientific machine learning and flow simulation}

Alongside the rapid development of neural architectures, standardized benchmarks have become essential for evaluating scientific machine learning methods. General-purpose PDE benchmarks, such as PDEBench \cite{takamoto2022pdebench} and PINNacle \cite{hao2023pinnacle}, provide valuable testbeds for assessing neural solvers across a range of partial differential equations. Fluid-oriented benchmarks, including CFDBench \cite{luo2023cfdbench}, FlowBench \cite{tali2024flowbench}, and FD-Bench \cite{wang2025fdbench}, further extend the evaluation landscape toward flow simulation over complex geometries and varying physical parameters. These efforts have played an important role in promoting reproducible comparison and accelerating progress in data-driven fluid dynamics.

Nevertheless, most existing benchmarks remain primarily focused on continuum-level flow problems. Their governing equations are usually based on the Navier-Stokes equations or other hydrodynamic PDEs, and the corresponding datasets often involve moderate flow conditions, smooth field structures, and standard macroscopic variables such as density, velocity, pressure, and temperature. While such settings are valuable for evaluating neural representations in conventional CFD tasks, they do not fully cover the physical complexity of non-equilibrium flow transport. In rarefied and hypersonic regimes, models must deal with Knudsen layers, velocity-distribution distortion, high-order stress and heat-flux moments, strong shock discontinuities, and multi-scale kinetic-to-continuum behavior.

Another challenge concerns the comparability of model evaluations. Existing studies often differ in parameter budgets, grid resolutions, preprocessing pipelines, train-test splits, optimization strategies, and levels of hyperparameter tuning. 
As a result, reported performance differences may be conflated with increased model capacity, higher data resolution, longer training, or more aggressive implementation-specific optimization. In such cases, it becomes difficult to determine whether a model performs better because of a more suitable architectural inductive bias or simply because it benefits from larger parameter counts and greater computational effort. This issue becomes particularly important in non-equilibrium flows, where model behavior is highly sensitive to spatial representation, normalization strategy, and the treatment of high-frequency features.

TransportBench is designed to complement existing benchmarks by focusing specifically on non-equilibrium flow transport. As summarized in Table \ref{tab:benchmark_comparison}, it differs from prior datasets in two main aspects. First, it includes high-fidelity non-equilibrium flow data generated from gas-kinetic simulations, enabling evaluation beyond standard Navier-Stokes-level benchmarks. Second, it incorporates high-order non-equilibrium quantities, such as stress tensor, heat flux, and particle distribution function, rather than only macroscopic conservative variables. In addition, it adopts unified preprocessing, training protocols, parameter budgets, and evaluation metrics to enable fair and manageable comparison across representative scientific machine learning architectures. Through this design, TransportBench serves both as a reusable dataset for non-equilibrium flow transport and as a standardized benchmark for assessing model performance across distinct flow regimes.

\begin{table}[htbp]
    \centering
    \caption{Comparison of TransportBench with existing scientific machine learning and fluid dynamics benchmarks in terms of physical fidelity, generalization capability, and evaluation difficulty. “Non-equilibrium effects” indicates whether the benchmark includes transport phenomena beyond the standard continuum-flow level. “High-order statistics” denotes prediction targets such as the stress tensor and heat flux. "Kinetic" denotes the Boltzmann equation and related physics.} 
    \label{tab:benchmark_comparison}
    \renewcommand{\arraystretch}{1.3} 
    \resizebox{\textwidth}{!}{
    \begin{tabular}{@{} l c c c c c c c @{}}
        \toprule
        & \multicolumn{3}{c}{\textbf{Physical Fidelity}} & \multicolumn{2}{c}{\textbf{Generalization Capability}} & \multicolumn{2}{c}{\textbf{Evaluation Difficulty}} \\
        \cmidrule(lr){2-4} \cmidrule(lr){5-6} \cmidrule(lr){7-8}
        \textbf{Benchmark} & 
        \begin{tabular}{@{}c@{}}Governing \\ equations\end{tabular} & 
        \begin{tabular}{@{}c@{}}Non-equilibrium \\ effects\end{tabular} & 
        \begin{tabular}{@{}c@{}}High-order \\ statistics\end{tabular} & 
        \begin{tabular}{@{}c@{}}Geometries \\ \end{tabular} & 
        \begin{tabular}{@{}c@{}}Physics \\ \end{tabular} & 
        \begin{tabular}{@{}c@{}}Discontinuities \\ \& shocks\end{tabular} & 
        \begin{tabular}{@{}c@{}}Sparse grids \\ \& few-shot\end{tabular} \\
        \midrule
        PDEBench \cite{takamoto2022pdebench} & Continuum (NS) & $\boldsymbol{\times}$ & $\boldsymbol{\times}$ & $\boldsymbol{\times}$ & \checkmark & $\boldsymbol{\times}$ & $\boldsymbol{\times}$ \\
        PINNacle \cite{hao2023pinnacle}      & Continuum (mixed)     & $\boldsymbol{\times}$ & $\boldsymbol{\times}$ & $\boldsymbol{\times}$ & $\boldsymbol{\times}$ & $\boldsymbol{\times}$ & $\boldsymbol{\times}$ \\
        CFDBench \cite{luo2023cfdbench}      & Continuum (NS) & $\boldsymbol{\times}$ & $\boldsymbol{\times}$ & \checkmark & \checkmark & $\boldsymbol{\times}$ & $\boldsymbol{\times}$ \\
        FlowBench \cite{tali2024flowbench}   & Continuum (NS) & $\boldsymbol{\times}$ & $\boldsymbol{\times}$ & \checkmark & \checkmark & $\boldsymbol{\times}$ & $\boldsymbol{\times}$ \\
        FD-Bench \cite{wang2025fdbench}      & Continuum (NS) & $\boldsymbol{\times}$ & $\boldsymbol{\times}$ & \checkmark & \checkmark & $\boldsymbol{\times}$ & $\boldsymbol{\times}$ \\
        \midrule
        \textbf{TransportBench} & \textbf{Kinetic} & \textbf{\checkmark} & \textbf{\checkmark} & \textbf{\checkmark} & \textbf{\checkmark} & \textbf{\checkmark} & \textbf{\checkmark} \\
        \bottomrule
    \end{tabular}
    }
\end{table}

\section{Dataset and Benchmark Design}
\label{sec:transportbench}


\subsection{Design principles}
\label{sec:principles}

TransportBench is designed with two closely connected objectives: to provide physically representative datasets for non-equilibrium flow transport and to establish controlled settings for benchmarking scientific machine learning models. The first objective concerns dataset construction. To this end, it covers a broad range of flow regimes, including rarefied and continuum fluids, low-speed and hypersonic conditions, inert and reacting gases, and both translational and internal-energy non-equilibrium effects. These regimes are selected to expose learning models to physical phenomena that are usually absent or underrepresented in standard continuum-flow benchmarks, such as Knudsen layers, velocity-distribution distortion, stress and heat-flux moments, thermochemical non-equilibrium, and shock-shock and shock-boundary interactions.

The second objective concerns model evaluation. TransportBench includes tasks that probe distinct learning challenges, including geometric generalization, physical-parameter generalization, high-order moment prediction, and shock-dominated field reconstruction. It also adopts unified preprocessing, training protocols, parameter budgets, and evaluation metrics, so that model comparisons are less affected by differences in data handling, model size, or implementation-specific training choices.

A particular design consideration is the representation of sharp physical structures. Numerical analyses show that neural networks tend to learn smooth, low-frequency components more easily than localized high-frequency variations \cite{jacot2018neural,rahaman2019spectral}. 
In shock-dominated and multi-scale flows, this tendency may lead to smeared discontinuities or inaccurate Rankine-Hugoniot values. 
A further question concerns the role of explicit high-frequency feature injection \cite{tancik2020fourier}. Such techniques are often introduced to mitigate spectral bias and improve the representation of sharp spatial variations. However, their effect may depend strongly on the underlying architecture. For models with different coordinate, convolutional, spectral, or integral representations, high-frequency injection may either improve shock resolution or introduce oscillatory artifacts, noisy background structures, or frequency-aliasing effects. TransportBench therefore includes shock-dominated configurations not only as difficult prediction tasks, but also as diagnostic settings for examining when high-frequency feature injection helps, when it fails, and how its effect interacts with architectural inductive bias.

Following these principles, the benchmark is constructed from four representative flow scenarios, which are introduced in detail below.

\subsection{Unified physical and learning formulation}
\label{sec:physical learning}

Non-equilibrium fluid transport can be described from a unified statistical mechanics perspective, where the state of a gas is characterized by distribution functions over particle velocity and internal states.
For a gas-dynamic system that may contain multiple species and internal energy modes, a generalized Boltzmann equation can be written as
\begin{equation}
    \frac{\partial f_{s,l}}{\partial t} + \mathbf{v} \cdot \nabla_{\mathbf{x}} f_{s,l} = \mathcal{Q}_{s,l}^{el} + \mathcal{Q}_{s,l}^{inel} + \mathcal{Q}_{s,l}^{chem},
    \label{eq:boltz}
\end{equation}
where $f_{s,l}(\mathbf x,\mathbf v,t)$ denotes the mass density distribution function at the $s$-th species in the $l$-th internal state.
The terms on the right-hand side represent elastic collisions, inelastic internal-energy transitions, and chemical production or loss processes, respectively.
This formulation provides a common physical basis for non-equilibrium transport, including rarefied and high-temperature gas dynamics.

For a monatomic gas without chemical reactions, Eq.(\ref{eq:boltz}) reduces to the classical Boltzmann equation,
\begin{equation}
    \frac{\partial f}{\partial t} + \mathbf{v} \cdot \nabla_{\mathbf{x}} f = \mathcal{Q}^{el}(f).
\end{equation}
Based on the distribution function, macroscopic flow variables can be obtained by taking moments over the velocity space, i.e.,
\begin{equation}
    \rho=\int_{\mathbb R^3}fd\mathbf v,\quad \rho \mathbf u=\int_{\mathbb R^3}\mathbf vfd\mathbf v,\quad \rho E=\int_{\mathbb R^3}\frac{1}{2}|\mathbf v|^2 fd\mathbf v.
\end{equation}
We use $\mathbf{U}(\mathbf{x}, t) = [\rho, \rho \mathbf u, \rho E]^\top$ to denote the conservative variables.
Taking moments of the Boltzmann equation yields macroscopic transport equations,
\begin{equation}
    \begin{aligned}
        &\frac{\partial \rho}{\partial t}+\nabla_{\mathbf{x}}\cdot (\rho \mathbf u)=0, \\
        &\frac{\partial \rho \mathbf u}{\partial t}+\nabla_{\mathbf{x}}\cdot (\rho \mathbf u\otimes\mathbf u+\mathbf P)=0, \\
        &\frac{\partial \rho E}{\partial t}+\nabla_{\mathbf{x}}\cdot (\rho E\mathbf u+\mathbf P\cdot\mathbf u+\mathbf q)=0, \\
    \end{aligned}
\end{equation}
where the total pressure tensor and heat flux are given by
\begin{equation}
    \mathbf P=\int_{\mathbb R^3} \mathbf c \otimes \mathbf c f d\mathbf v,\quad \mathbf q=\int_{\mathbb R^3} \frac{1}{2} |\mathbf c|^2 \mathbf cfd\mathbf v,
\end{equation}
where $\mathbf c=\mathbf v-\mathbf u$ is the peculiar velocity.
For non-equilibrium flows, higher-order moments, including stress tensor and heat flux, are essential to characterize deviations from local equilibrium, and they are explicitly included in selected TransportBench tasks via numerical simulations of the Boltzmann equation and DSMC.

With the constitutive relations incorporated, solvable macroscopic equations can be obtained.
This can be helpful when directly solving the Boltzmann equation is overwhelmingly costly.
For high-temperature reacting flows, the vibrational state-specific conservation laws can be derived as an asymptotic limit of Eq.(\ref{eq:boltz}) and written as
\begin{equation}
    \begin{aligned}
        &\frac{\partial \rho_{s,l}}{\partial t} + \nabla_{\mathbf{x}} \cdot (\rho_{s,l}\mathbf{u}) = \nabla_{\mathbf{x}} \cdot (\rho D_s \nabla_{\mathbf{x}} \varphi_{s,l}) + S_{s,l}, \\
        &\frac{\partial \rho \mathbf{u}}{\partial t} + \nabla_{\mathbf{x}} \cdot (\rho \mathbf{u} \otimes \mathbf{u} + \mathbf{P}) = 0, \\
        &\frac{\partial \rho E}{\partial t} + \nabla_{\mathbf{x}} \cdot (\rho E \mathbf{u} + \mathbf{P} \cdot \mathbf{u} + \mathbf{q}) = \nabla_{\mathbf{x}} \cdot \left( \rho \sum_{s=1}^{N_s} D_s \sum_{l=1}^{V_s} h_{s,l} \nabla_{\mathbf{x}} \varphi_{s,l} \right), \\
    \end{aligned}
\end{equation}
where the total pressure tensor and heat flux are given by
\begin{equation}
    \mathbf{P} = p\mathbf{I} - \mu \left( \nabla_{\mathbf{x}}\mathbf{u} + (\nabla_{\mathbf{x}}\mathbf{u})^T \right) + \frac{2}{3}\mu(\nabla_{\mathbf{x}} \cdot \mathbf{u})\mathbf{I}, \quad \mathbf{q} = -\eta \nabla_{\mathbf{x}} T,
\end{equation}
where $\rho_{s,l}$, $\varphi_{s,l}$, $h_{s,l}$ and $S_{s,l}$ denote the density, mass fraction, enthalpy, and source term of the $s$-th species in the $l$-th internal state, respectively.
Besides, $N_s$ represents the total number of chemical species, and $V_s$ is the number of vibrational states associated with the $s$-th species, $D_s$ is the diffusion coefficient of the $s$-th species, $\eta$ is the thermal conductivity associated with translational-rotational energy, and $\mu$ is the shear viscosity.

From the learning perspective, all tasks are formulated in a unified way as input-output mappings. Let $\mathcal A$ denote the input operating condition, which may include geometry descriptions and freestream parameters. Let $\mathcal U$ denote the target solution field, which may include macroscopic quantities such as density, velocity, pressure, and temperature, as well as non-equilibrium quantities such as stress tensor, heat flux, and particle distribution function. The learning problem is written as
\begin{equation}
    \mathcal G:\  \mathcal A\mapsto \mathcal U.
\end{equation}
A scientific machine learning model $\mathcal G_\theta$ is trained to approximate this mapping using a reconstruction loss,
\begin{equation}
    \label{eq:pil}
    \mathcal{L}_{\text{rec}} = \left\| \mathcal{G}_\theta(\mathcal A)-\mathcal U \right\|_2^2.
\end{equation}
When physical constraints are imposed, the training objective can be augmented as
\begin{equation}
    \mathcal{L} = \mathcal{L}_{\text{rec}} + \lambda_{\text{phys}} \mathcal{L}_{\text{phys}},
\end{equation}
where $\mathcal{L}_{\text{phys}}$ may represent residuals of kinetic equations, conservation laws, or other task-specific constraints described above, and $\lambda_{\text{phys}}$ is the corresponding weighting coefficient.
This formulation accommodates both purely data-driven baselines and physics-informed extensions, while maintaining a unified view of TransportBench as a benchmark for learning non-equilibrium flow transport across multiple modeling levels.

\subsection{Physical scenarios}
\label{sec:scenarios}

\begin{figure}[htbp]
    \centering
    \includegraphics[width=\textwidth]{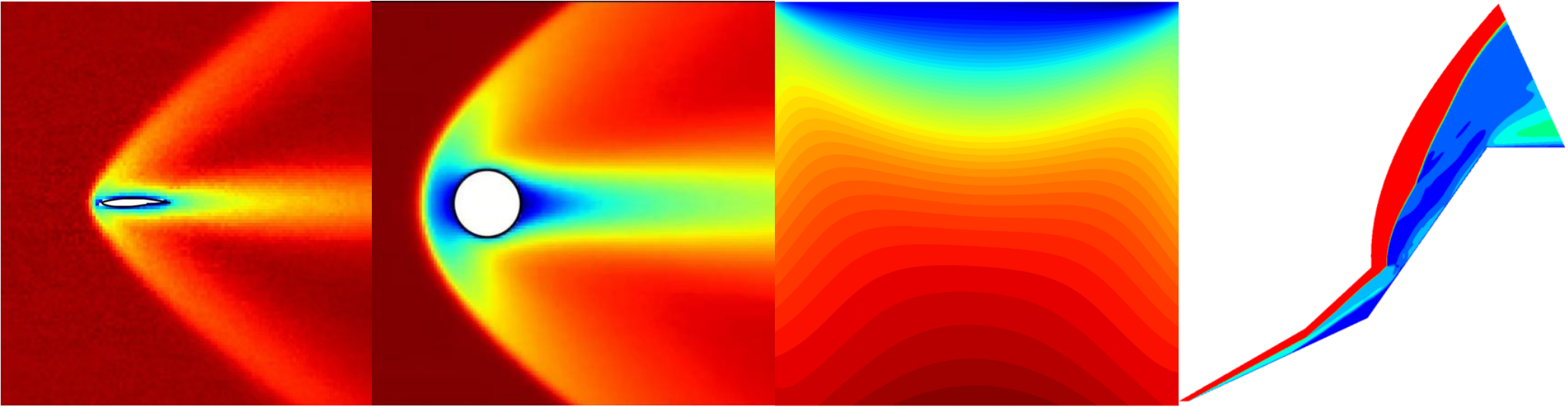} 
    \caption{\textbf{Representative ground-truth flow fields of the four physical scenarios in TransportBench.} From left to right: (a) rarefied airfoil flows with varying geometries for geometry-dependent predictions; (b) rarefied cylinder flow across a range of Mach and Knudsen numbers for operating-condition-dependent predictions; (c) lid-driven cavity flows for distribution-function and high-order moment predictions; and (d) high-enthalpy double-cone flows for shock-dominated hypersonic field reconstructions under thermochemical non-equilibrium. Together, these scenarios cover geometry variation, operating-condition variation, kinetic-moment prediction, and sharp-discontinuity representation in non-equilibrium fluid transport.}
    \label{fig:task_overview}
\end{figure}

Based on the above design principles and unified formulation, TransportBench contains four representative flow scenarios. They are selected to cover different levels of non-equilibrium gas transport and to expose neural network models to complementary learning challenges, including geometry-dependent prediction, parameter-dependent prediction, high-order kinetic-moment prediction, and shock-dominated field reconstruction. Figure \ref{fig:task_overview} shows the ground-truth flow fields for four representative tasks.


\vspace{0.5em}
\noindent\textbf{Task I: airfoil flow with geometry-dependent prediction} 

Flow over an airfoil is a classical problem in aerodynamics. In rarefied gas dynamics, the interaction between the incoming flow and the airfoil surface can generate complex structures, including bow shocks near the leading edge, boundary-layer development, and wake regions downstream of the profile.
When the airfoil geometry changes, the shock standoff distance, surface interaction, and wake morphology also vary. This makes the airfoil case suitable for evaluating whether neural models can learn geometry-dependent flow structures and predict flow fields for unseen airfoil shapes.

\begin{figure}[htbp]
    \centering
    \begin{subfigure}[b]{0.45\textwidth}
        \includegraphics[width=\textwidth]{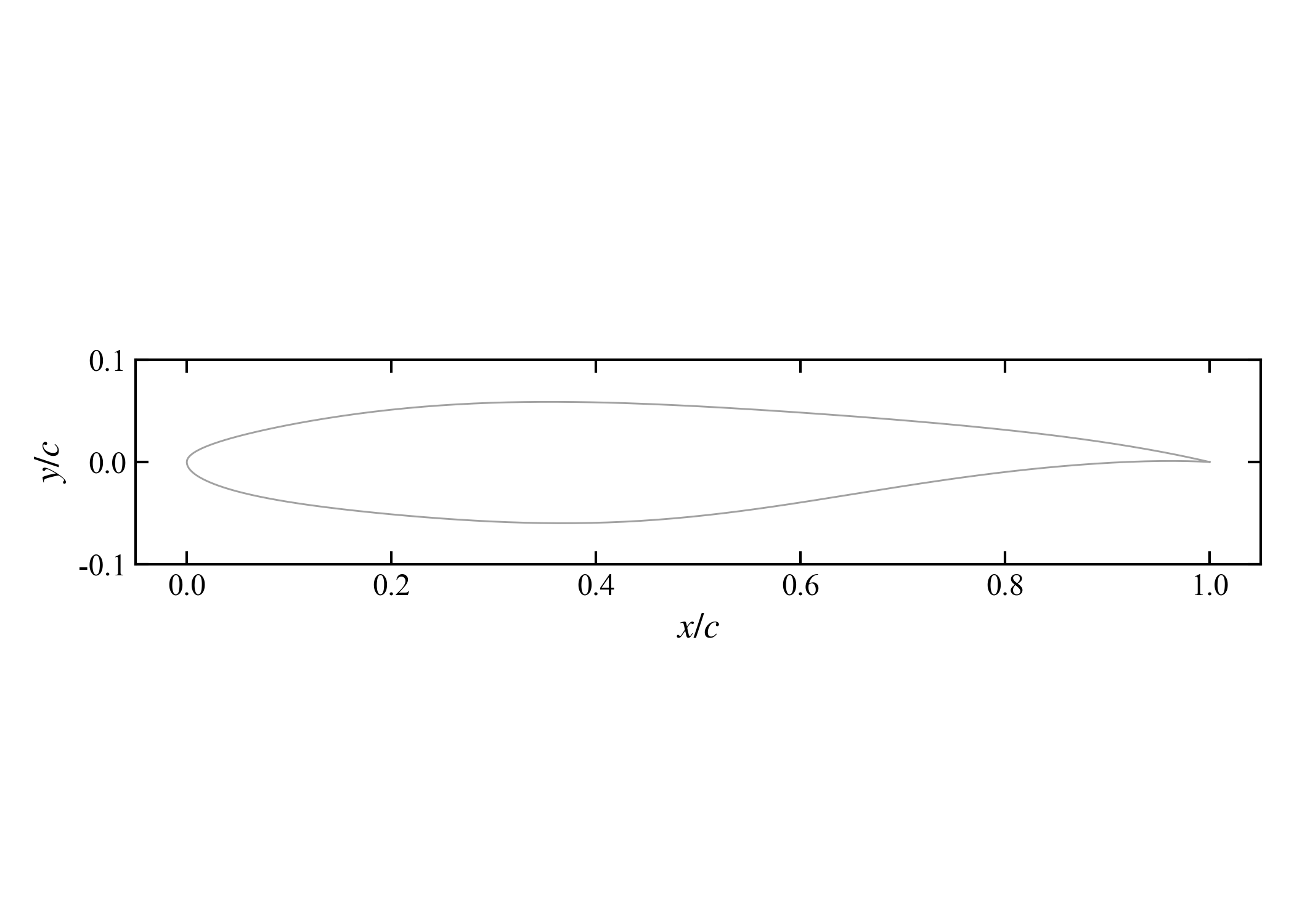}
        \caption{RAE2822 baseline airfoil}
    \end{subfigure}
    \hspace{0.01\linewidth}
    \begin{subfigure}[b]{0.45\textwidth}
        \includegraphics[width=\textwidth]{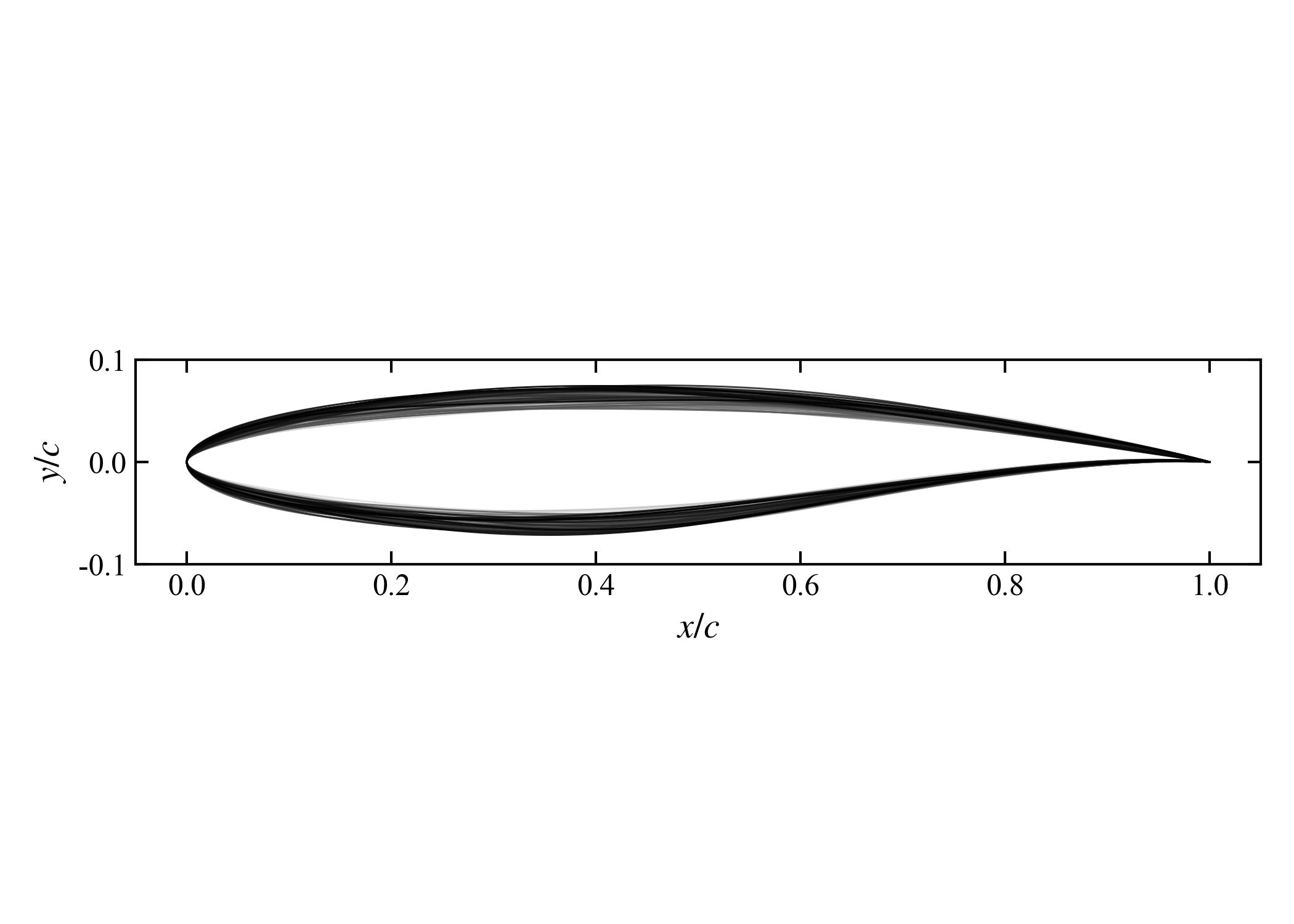}
        \caption{Airfoil samples via CST perturbation}
    \end{subfigure}
    \caption{Baseline RAE2822 airfoil and CST-generated geometric variations.}
    \label{Fig_RAE_GridAnalysis}
\end{figure}

The baseline geometry is the RAE2822 airfoil, parameterized using the Class-Shape Transformation (CST) method.
The upper and lower surfaces are represented by sixth-order Bernstein polynomials.
The corresponding CST coefficients of the baseline airfoil are
\begin{equation}
\begin{aligned}
    &\mathbf{b}_U = [0.1266,\ 0.1379,\ 0.1617,\ 0.1642,\ 0.2205,\ 0.1719,\ 0.2132], \\
    &\mathbf{b}_L = [-0.1289,\ -0.1426,\ -0.1311,\ -0.2615,\ -0.0253,\ -0.1145,\ 0.0727],
\end{aligned}
\end{equation}
where $\mathbf{b}_U$ and $\mathbf{b}_L$ denote the CST coefficients of the upper and lower surfaces, respectively.
To generate geometric variations, each coefficient is independently perturbed within $\pm 30\%$,
\begin{equation}
    b_i' = b_i \cdot (1 + r_i),\quad r_i \sim \mathscr{U}(-0.3, 0.3),
\end{equation}
where $\mathscr U$ denotes the uniform distribution.
A total of 100 shape-varied airfoil geometries are produced, as illustrated in Figure~\ref{Fig_RAE_GridAnalysis}.

The incoming flow consists of nitrogen and is set to the Mach number $\rm Ma=2.0$ and the Knudsen number $\rm Kn=0.026$, corresponding to the transition regime. The freestream temperature is $161\ \mathrm{K}$, the freestream velocity is $517.26\,\mathrm{m/s}$, and the wall temperature is fixed at $290\,\mathrm{K}$. A fully accommodating diffuse reflection model is imposed at the airfoil surface interaction. These physical parameters are kept fixed for all 100 samples so that the task isolates the effect of geometric variation. The resulting flow fields are mapped onto a Cartesian grid of $128 \times 128$.
In the benchmark evaluation, the train-test split is performed over the set of airfoil geometries, so that the task assesses both interpolation within the geometry distribution and prediction for unseen shapes.

\vspace{0.5em}
\noindent\textbf{Task II: cylinder flow with parameter-dependent prediction}

Flow around a circular cylinder is a canonical configuration for studying flow separation and wake dynamics.
In high-speed rarefied gas flow, the bow shock and downstream wake are strongly affected by compressibility and rarefaction effects.
The cylinder case is therefore designed to evaluate whether neural models can learn parameter-dependent changes in flow topology over a wide range of operating conditions, especially Mach and Knudsen numbers.


In this task, the cylinder geometry is fixed, while the operating conditions vary over a broad parameter space.
A total of 2,400 flow cases are generated.
The Knudsen number ranges from $\rm Kn=0.06$ to $\rm Kn=1.0$ with increments of 0.02, and the Mach number ranges from $\rm Ma=2.0$ to $\rm Ma=6.9$ with increments of 0.1.
Argon is used as the working gas. The freestream temperature is set to \( T_{\infty} = 273 \) K, and the wall temperature is fixed at \( T_w=273 \)K. The gas-surface interaction at the cylinder wall is modeled using fully diffuse reflection.

\begin{figure}[htbp]
    \centering
    \begin{subfigure}[b]{0.45\textwidth}
        \includegraphics[width=\textwidth]{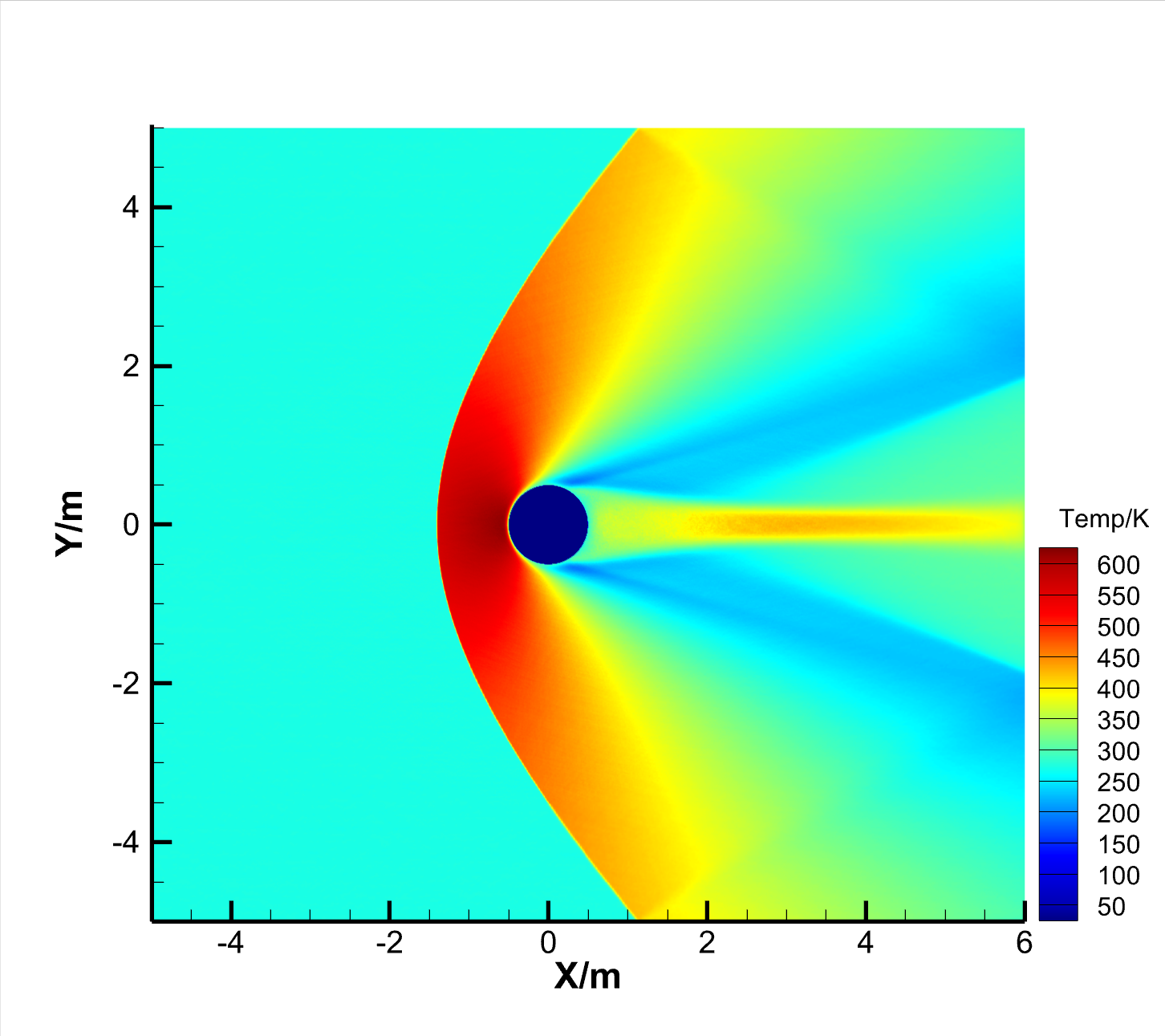}
        \caption{$\mathrm{Kn}=0.06, \mathrm{Ma}=2.0$}
        \label{fig:cyl_a}
    \end{subfigure}
    \hspace{0.02\linewidth} 
    \begin{subfigure}[b]{0.45\textwidth}
        \includegraphics[width=\textwidth]{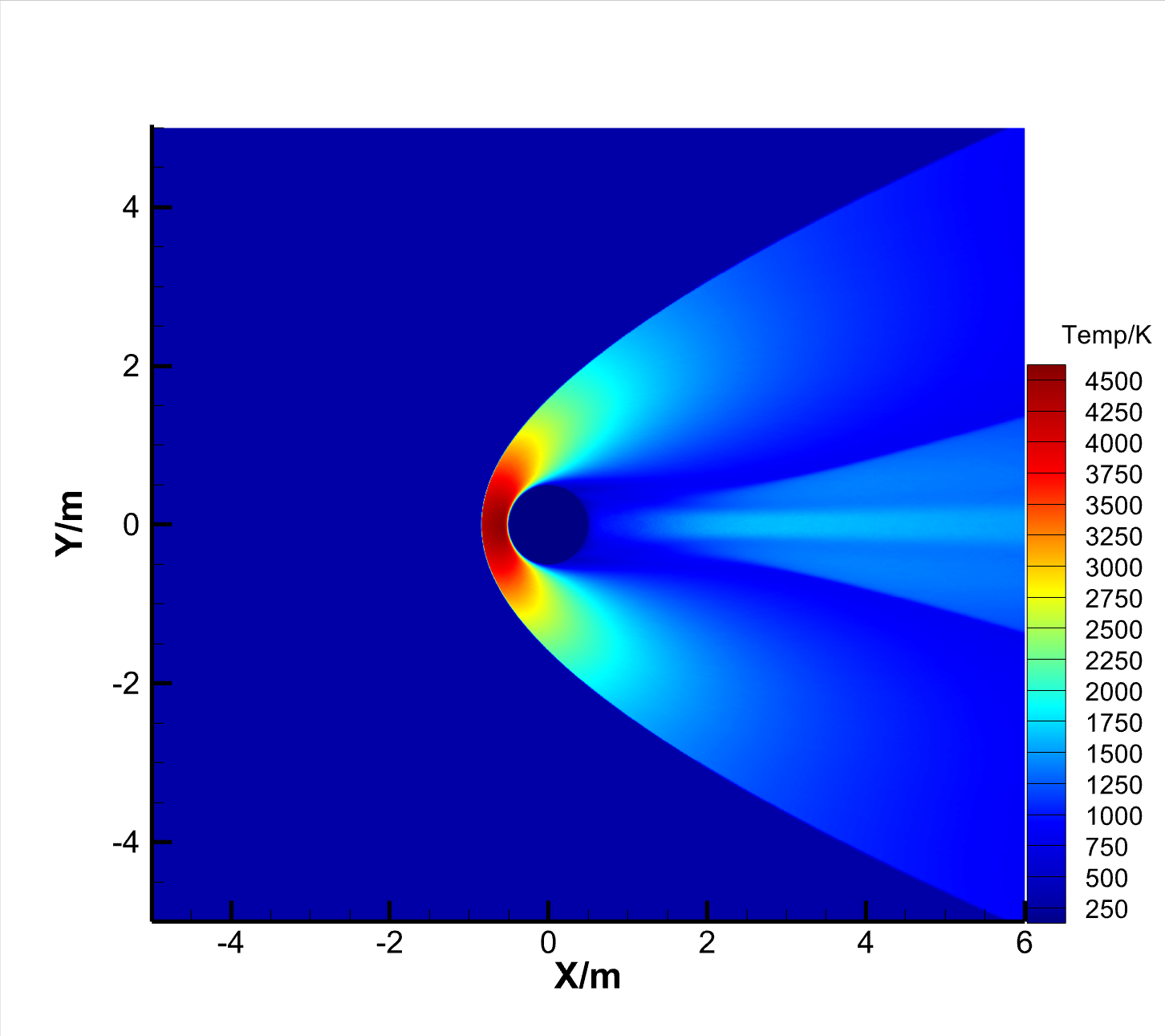}
        \caption{$\mathrm{Kn}=0.06, \mathrm{Ma}=6.9$}
        \label{fig:cyl_b}
    \end{subfigure}
    
    \vspace{0.5em} 
    
    \begin{subfigure}[b]{0.45\textwidth}
        \includegraphics[width=\textwidth]{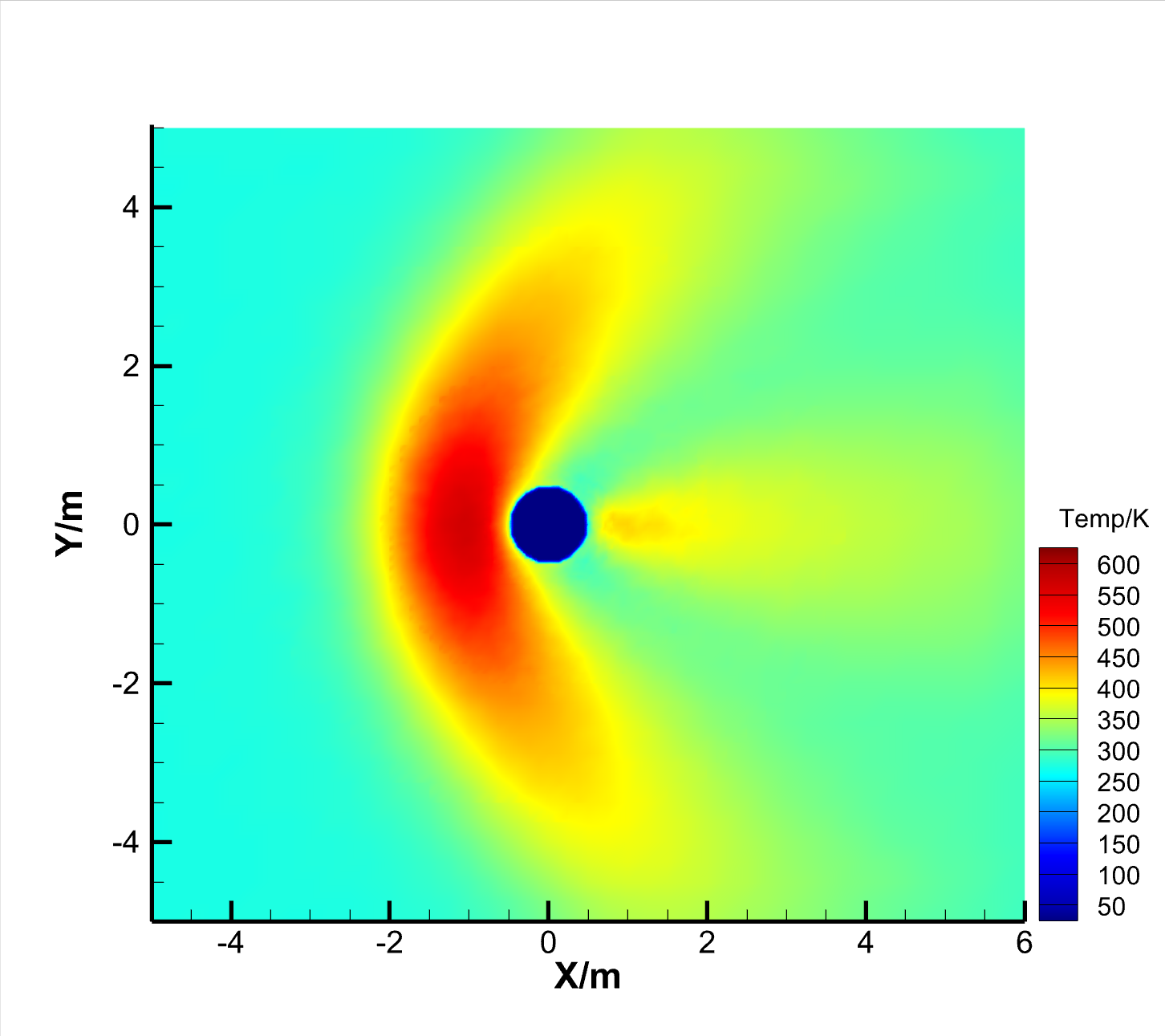}
        \caption{$\mathrm{Kn}=1.0, \mathrm{Ma}=2.0$}
        \label{fig:cyl_c}
    \end{subfigure}
    \hspace{0.02\linewidth} 
    \begin{subfigure}[b]{0.45\textwidth}
        \includegraphics[width=\textwidth]{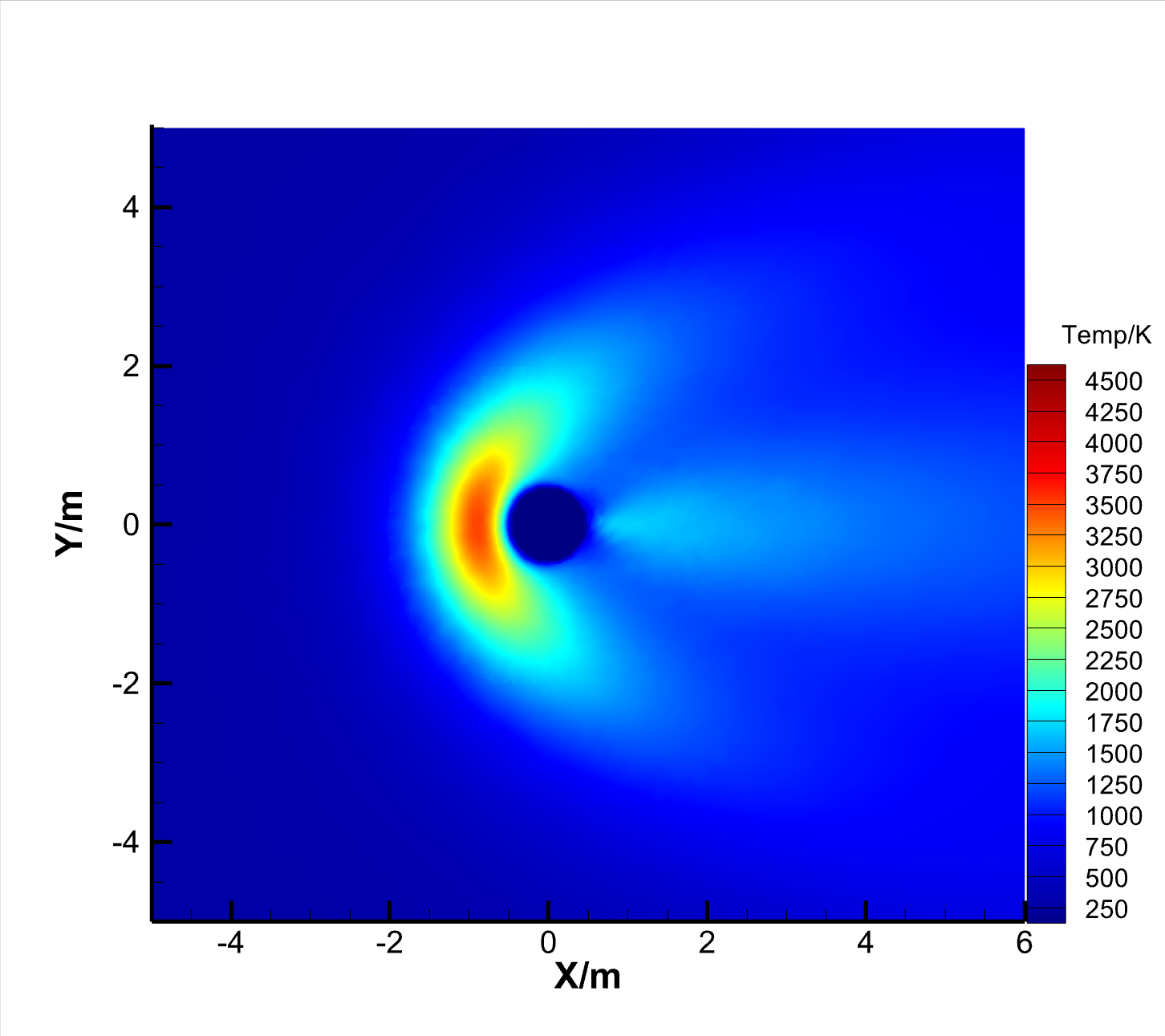}
        \caption{$\mathrm{Kn}=1.0, \mathrm{Ma}=6.9$}
        \label{fig:cyl_d}
    \end{subfigure}
    
    \caption{Representative temperature fields in the cylinder flow. The four cases correspond to different combinations of Mach and Knudsen numbers. At lower Knudsen numbers, clearer bow-shock and wake structures are observed, whereas at higher Knudsen numbers the reduced collision frequency leads to stronger rarefaction effects and more diffuse macroscopic flow structures. This comparison illustrates the freestream-condition-dependent variation of shock and wake topology.}
    \label{Fig_Cylinder_FlowField}
\end{figure}

Across this parameter space, the flow topology changes significantly.
At lower Knudsen numbers, the flow exhibits clearer bow-shock and wake structures. At higher Knudsen numbers, the frequency of intermolecular collisions decreases, the rarefaction effect becomes more pronounced, and the scale of macroscopic flow structures increases accordingly, as shown in Figure \ref{Fig_Cylinder_FlowField}. The computational domain extends 11 cylinder radii in the streamwise direction and 10 cylinder radii in the transverse direction. 
A multi-level grid refinement strategy is used to improve near-wall resolution. Depending on the boundary-layer thickness and rarefaction level, the number of unstructured cells varies across cases. The data are then mapped onto an elongated Cartesian grid of $128\times 192$ to capture the extended wake region and non-equilibrium transport.


This task focuses on predictions across varying operating conditions. It tests whether a neural architecture can learn how global flow structures vary with rarefaction and compressibility, rather than only interpolate between visually similar flow fields. The test set contains unseen combinations of Mach and Knudsen numbers, enabling evaluation of parameter-space prediction and generalization under controlled conditions.

\vspace{0.5em}
\noindent\textbf{Task III: lid-driven cavity flow with high-order kinetic prediction}

While Tasks I and II focus mainly on macroscopic flow-field prediction under geometry or parameter variation, Task III is designed to probe the micro-macro connection in non-equilibrium fluid transport.
The lid-driven cavity case is selected as its simple geometry and confined domain help isolate the gas-kinetic challenge associated with the Boltzmann equation. In the rarefied flow regime, the particle distribution function can deviate substantially from the local Maxwellian, and non-equilibrium transport phenomena emerge.
Therefore, predicting only low-order macroscopic variables is insufficient for fully characterizing the flow.

The physical domain is a two-dimensional unit square, $x, y \in [0, 1]$. A total of 99 simulation cases are generated, with the Knudsen number varying from Kn=0.02 to Kn=1.0 in increments of 0.01. This range covers multiple flow regimes and captures the evolution from near-equilibrium to strongly non-equilibrium states. Argon is used as the working gas. The boundary conditions consist of isothermal stationary walls at $T_w=273\ \rm K$ and a top lid moving at a constant velocity $u_w=50.55\ \rm m/s$.

The output of this task includes both macroscopic flow quantities and particle distribution functions.
The particle distribution function is resolved on a discrete velocity grid, and the macroscopic quantities are obtained by taking moments over velocity space.
Higher-order non-equilibrium statistics, including the stress tensor $\mathbf P$ and heat flux $\mathbf q$, are of particular importance because they encode the distortion of the velocity distribution function away from local equilibrium at a macroscopic level.
Compared with low-order macroscopic variables, these quantities provide a more sensitive measure of non-equilibrium transport.

\begin{figure}[htbp]
    \centering
    \begin{subfigure}[b]{0.48\textwidth}
        \includegraphics[width=\textwidth]{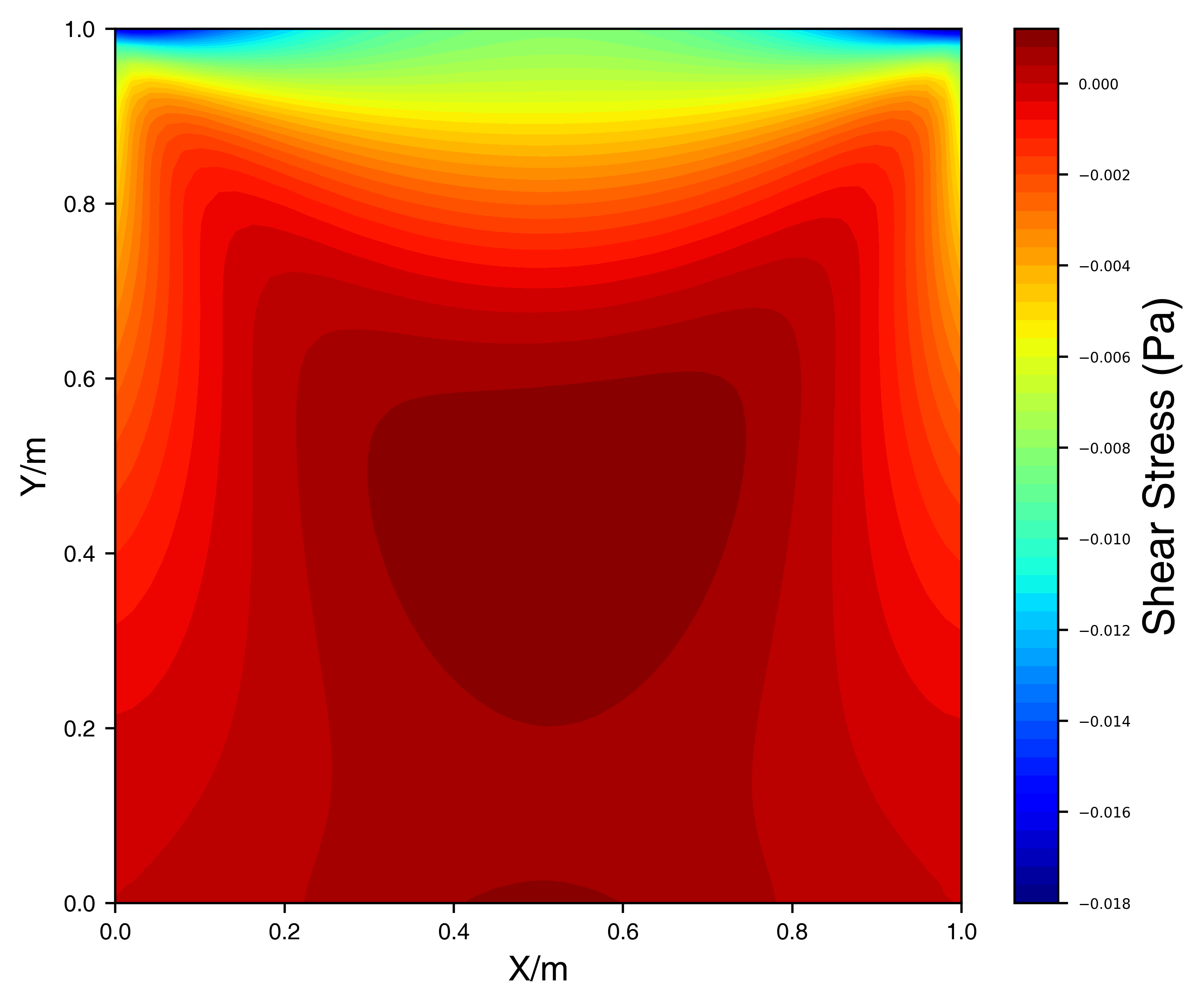}
        \caption{Shear stress $P_{xy}$ at $\mathrm{Kn}=0.02$}
    \end{subfigure}
    \hfill
    \begin{subfigure}[b]{0.48\textwidth}
        \includegraphics[width=\textwidth]{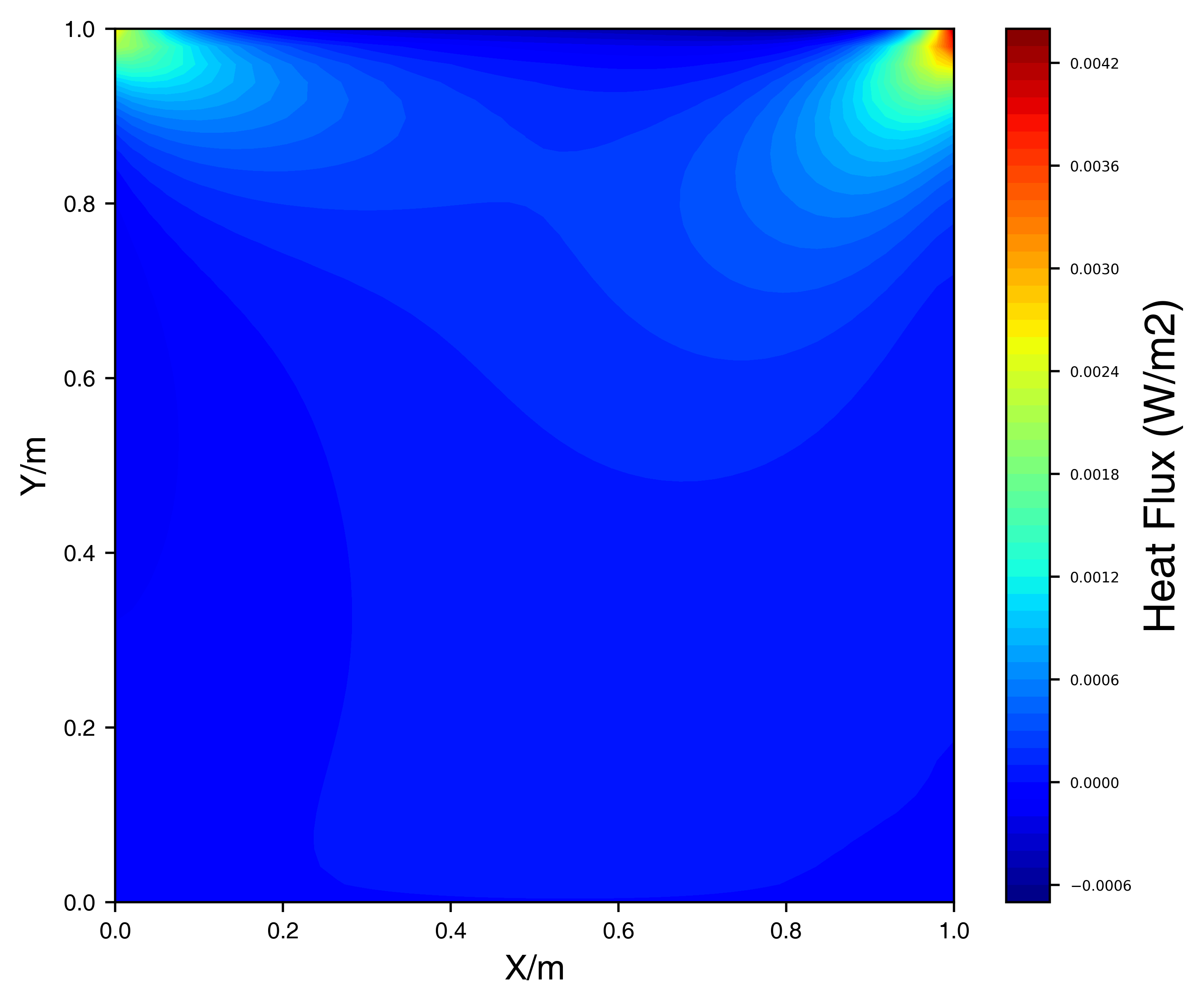}
        \caption{Heat flux $q_x$ at $\mathrm{Kn}=0.02$}
    \end{subfigure}
    
    \vspace{0.5em} 
    
    \begin{subfigure}[b]{0.48\textwidth}
        \includegraphics[width=\textwidth]{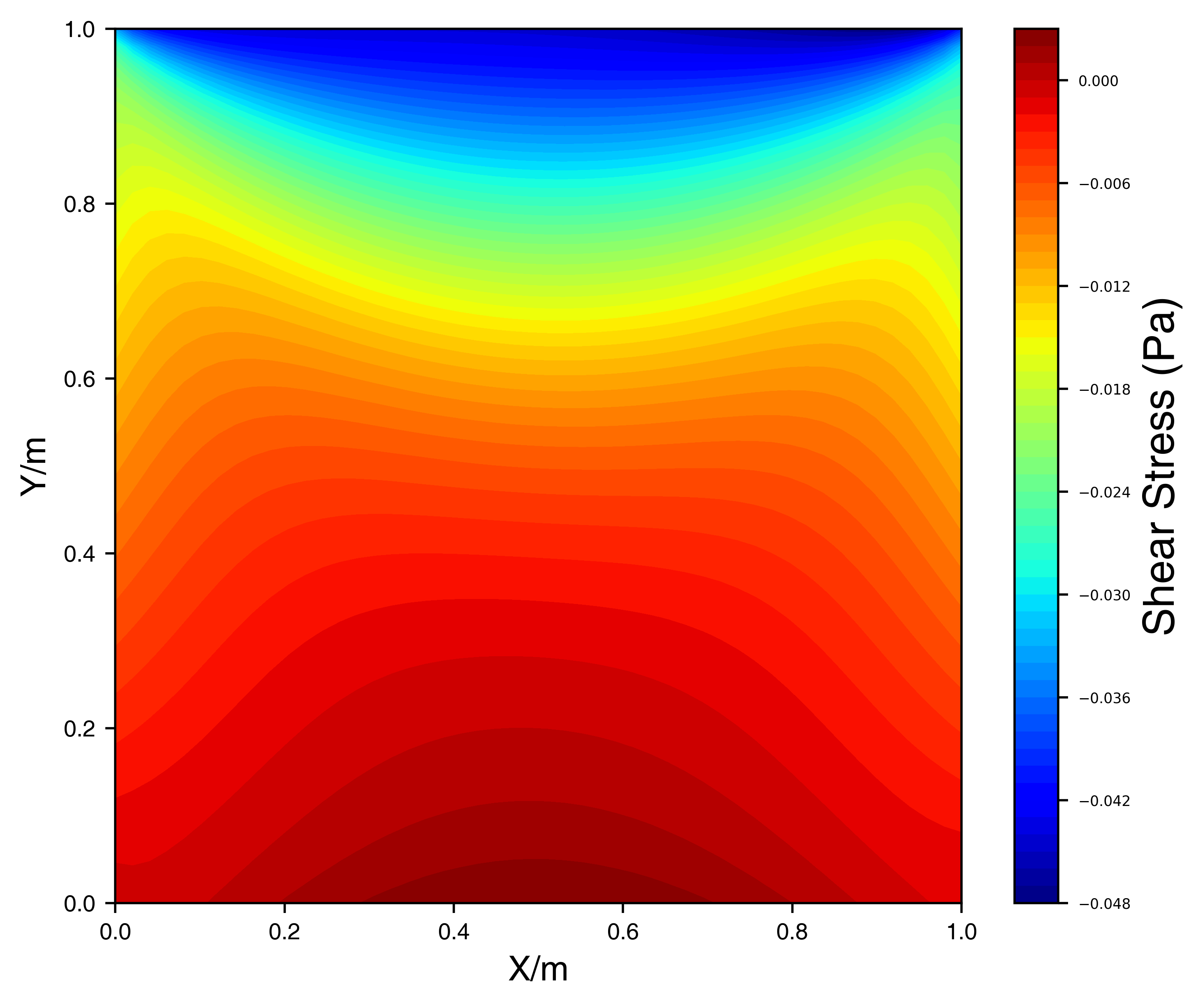}
        \caption{Shear stress $P_{xy}$ at $\mathrm{Kn}=1.0$}
    \end{subfigure}
    \hfill
    \begin{subfigure}[b]{0.48\textwidth}
        \includegraphics[width=\textwidth]{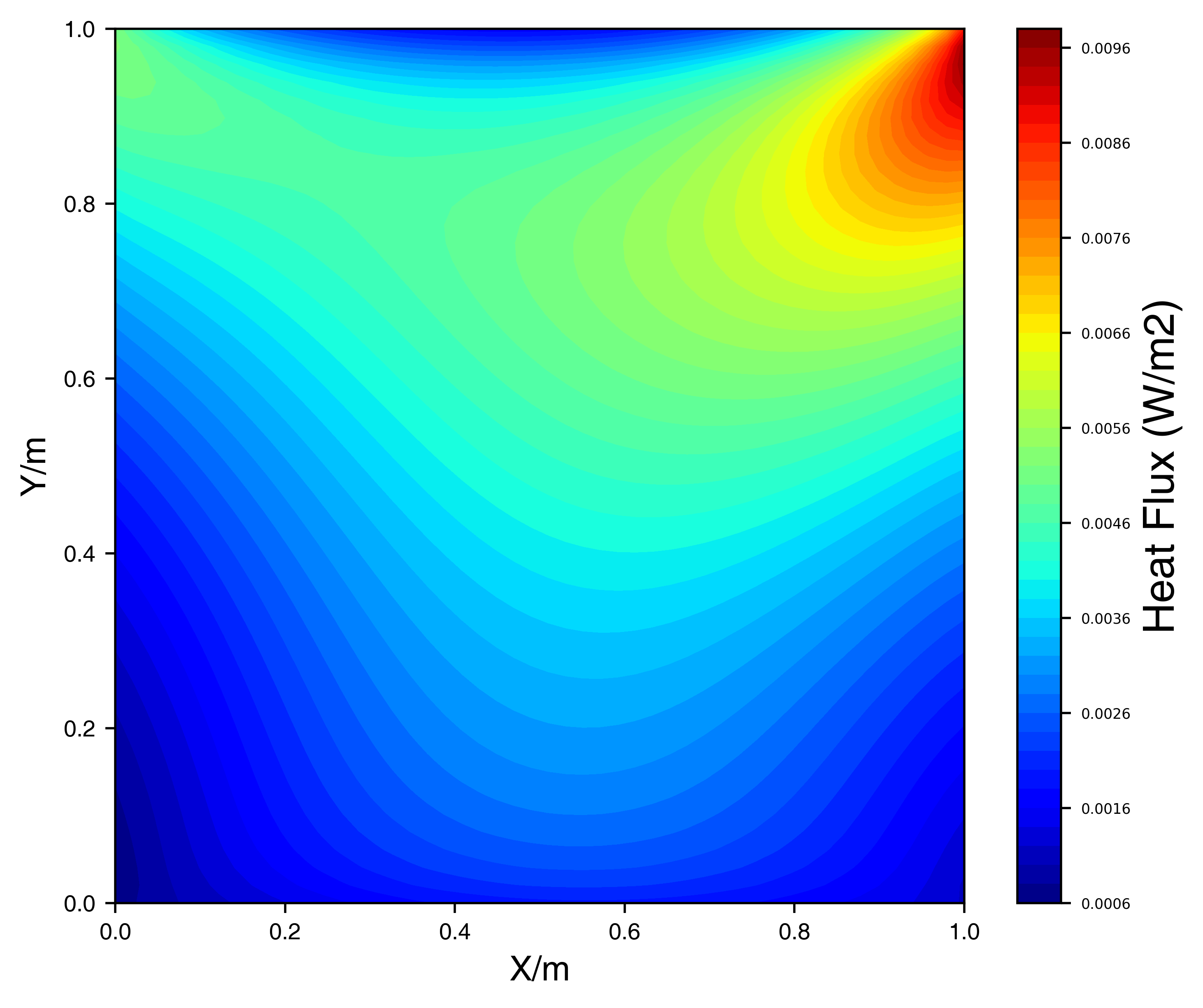}
        \caption{Heat flux $q_x$ at $\mathrm{Kn}=1.0$}
    \end{subfigure}
    
    \caption{Representative non-equilibrium moment fields in the lid-driven cavity flow. The figure shows the shear-stress component $P_{xy}$
 and heat-flux component $q_x$. The top row corresponds to the near-equilibrium regime at Kn=0.02, where the fields are relatively smooth; the bottom row corresponds to Kn=1.0, where stronger rarefaction effects lead to more pronounced gradients and non-equilibrium moment structures.}
    \label{Fig_Cavity_NonEq}
\end{figure}


Figure \ref{Fig_Cavity_NonEq} shows representative stress and heat-flux fields at different Knudsen numbers.
In the near-equilibrium regime at $\mathrm{Kn}=0.02$, these fields remain relatively smooth, whereas at higher Knudsen numbers they exhibit stronger non-equilibrium structures and pronounced gradients.
This task therefore evaluates whether neural network models can serve as a micro-macro bridge. An effective model should not only reproduce macroscopic fields, but also capture deviations of the particle distribution function from equilibrium.
Since the Knudsen number is varied across the dataset, this task also provides a controlled setting for testing prediction across different degrees of rarefaction.

\vspace{0.5em}
\noindent\textbf{Task IV: double-cone flow with shock-dominated field reconstruction}

This task considers high-enthalpy nitrogen flow over a $25^\circ$-$55^\circ$ double-cone configuration based on the CUBRC LENS-I experiments \cite{nompelis2003computational}.
Unlike Tasks I-III, which are dominated by translational non-equilibrium effects, this case is strongly influenced by thermochemical non-equilibrium.
Its physical complexity arises from detailed state-to-state (StS) kinetics for internal-energy relaxation and finite-rate processes. 
In the StS model, $\mathrm{N}_2$ is represented using 55 discrete vibrational energy levels. The state-specific mass production rates are evaluated by accounting for vibrational-translational (V-T) energy exchange and dissociation-recombination (D-R) processes associated with $\mathrm{N}_2$-$\mathrm{N}_2$ and $\mathrm{N}_2$-$\mathrm{N}$ collisions \cite{wang2023high}.
At the same time, the flow contains strong shock discontinuities and shock-boundary-layer interactions, making it a challenging case for neural field representation.

\begin{figure}[htbp]
    \centering
    \includegraphics[width=0.48\textwidth]{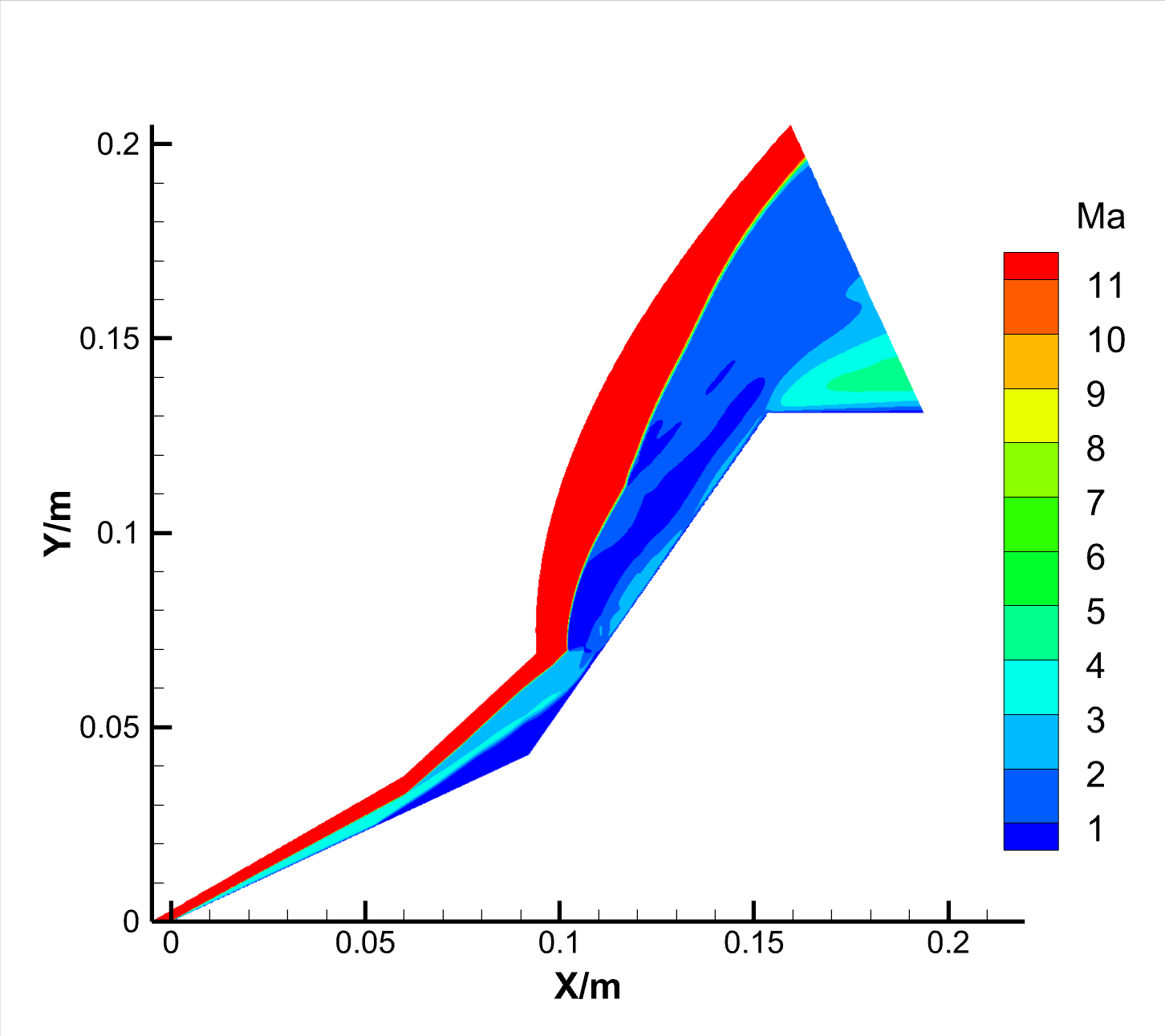} 
    \caption{Representative Mach-number contour for the high-enthalpy double-cone flow at $\mathrm{Ma}=11.628$ and $\mathrm{Re}=341500$.}
    \label{fig:doublecone_snapshot1}
\end{figure}

An axisymmetric computational grid of $1024 \times 512$ nodes is employed in the axial and radial directions, respectively. The first near-wall grid spacing is set to $1 \times 10^{-7}\,\mathrm{m}$ so that the cell Reynolds number is on the order of unity. This grid resolution has been verified to be grid-independent for simulations of high-enthalpy double cone flows \cite{wang2023high}.
The freestream Mach number ranges from Ma=6.0 to Ma=17.6, and the Reynolds number ranges from $1.4 \times 10^5$ to $2.3 \times 10^6$. As the Mach number increases, thermochemical non-equilibrium effects become more pronounced, and the vibrational energy distribution deviates more strongly from the equilibrium Boltzmann distribution. 
A representative Mach-number field is shown in Figure \ref{fig:doublecone_snapshot1}.

For learning-based evaluation, the data are mapped onto a sparse and highly anisotropic representation of $17 \times 384$. This setting is intentionally challenging. The dataset contains 51 high-fidelity snapshots, with 45 samples used for training and 6 samples used for testing. Meanwhile, the target fields contain localized shock peaks, steep gradients, and large property variations across narrow spatial regions. Therefore, this task is not merely a high-enthalpy flow prediction problem, but also a diagnostic test for the representational behavior of neural architectures under sparse and shock-dominated conditions.


This design directly reflects the frequency-representation issue discussed in the benchmark principles. Neural models that favor smooth, low-frequency mappings may capture the global flow topology but smear shock peaks and underestimate localized extrema. In contrast, models equipped with explicit high-frequency feature injection may better represent sharp structures in some cases, but may also introduce oscillatory artifacts, nonphysical noise, or frequency-aliasing effects. By including the double-cone configuration, this task provides a controlled setting for examining how architectural inductive bias, spectral bias, and high-frequency feature representations affect shock-dominated flow prediction.

Together, the four tasks form a benchmark suite that spans geometry-dependent prediction, parameter-dependent prediction, high-order kinetic-statistics prediction, and shock-dominated hypersonic field reconstruction. This organization allows TransportBench to evaluate both the predictive accuracy and the physical robustness of scientific machine learning models across distinct non-equilibrium flow regimes.

\section{Dataset Generation and Preprocessing}
\label{sec:dataset_gen}


\subsection{High-fidelity data generation}

Generating reliable reference data for non-equilibrium flow transport requires numerical methods that are consistent with the physical regime.
In TransportBench, the four tasks are therefore generated using different high-fidelity solvers and modeling levels.
The rarefied external flows are simulated using particle-based kinetic methods, the cavity flow is generated using a deterministic discrete velocity method for the Boltzmann equation, and the high-enthalpy double-cone task is computed using a continuum solver coupled with state-to-state thermochemical non-equilibrium modeling.
The solver selection is consistent with the unified physical formulation described in Section \ref{sec:physical learning}.

For Tasks I and II, the reference data are generated using the direct simulation Monte Carlo (DSMC) method \cite{bird1994molecular}. DSMC is a well-established gas-kinetic particle method for rarefied gas dynamics and is particularly suitable for high-speed flows with strong non-equilibrium effects \cite{Boyd_Schwartzentruber_2017}. A typical DSMC simulation alternates between particle transport and stochastic intermolecular collisions within computational cells. After the flow reaches a statistically steady state, macroscopic quantities such as density, velocity, temperature, stress, and heat flux can be obtained by sampling and averaging particle information over sufficiently long time intervals.

The open-source SPARTA solver \cite{plimpton2019direct} is used for the DSMC simulations. The variable hard sphere (VHS) molecule model is adopted to describe intermolecular collisions, and the no-time-counter (NTC) algorithm is used for efficient collision computations \cite{bird1994molecular}. The VHS parameters for argon and nitrogen are listed in Table \ref{tab:vhs_params}. For nitrogen internal-energy relaxation, a constant-collision-number model is used, with the rotational and vibrational collision numbers set to 5 and 50, respectively. Gas-surface interaction is modeled using fully diffuse reflection with complete thermal accommodation to the wall temperature. No additional surface reaction or recombination process is considered. Hierarchical Cartesian grids with adaptive refinement near solid boundaries are used to resolve near-wall rarefaction effects and Knudsen-layer structures. To reduce the stochastic noise inherent in particle methods, time-averaged sampling is performed over sufficiently long steady-state intervals.

\begin{table}[htbp]
\centering
\caption{VHS model parameters for Argon and Nitrogen ($T_{\text{ref}} = 273$ K).}
\label{tab:vhs_params}
\begin{tabular*}{14cm}{l@{\extracolsep{\fill}}c@{\extracolsep{\fill}}c}
\hline
Species & $\omega$ & $d_{\text{ref}}$ (\AA) \\
\hline
Ar & 0.81 & 4.17 \\
N$_2$ & 0.74 & 4.17 \\
\hline
\end{tabular*}
\end{table}

For Task III, the lid-driven cavity dataset is generated by solving the Boltzmann-BGK equation using a discrete velocity method (DVM) \cite{xiao2020velocity,xiao2021kinetic}. Unlike DSMC, which relies on stochastic particle sampling, the DVM resolves the distribution function on a discrete velocity grid and evolves it deterministically. Macroscopic variables and high-order non-equilibrium moments are then obtained by numerical integration over the velocity space. In particular, the stress tensor and heat flux are computed directly from velocity moments of the distribution function. This setting provides a clean kinetic dataset for evaluating whether neural models can predict high-order non-equilibrium statistics associated with microscopic velocity-distribution distortion.

For Task IV, the high-enthalpy double-cone flow is computed using the in-house ARTIST-CFD solver \cite{wang2023accuracy}. The governing equations are the compressible Navier-Stokes equations coupled with a vibrational state-to-state model for thermochemical non-equilibrium. Inviscid fluxes are computed using a modified Steger-Warming upwind scheme combined with second-order MUSCL reconstruction and a minmod limiter, while viscous fluxes are discretized using a second-order central scheme. Time advancement is performed using a data-parallel line relaxation method. Species viscosities and thermal conductivities are evaluated using Blottner’s curve fitting and Eucken’s relation, respectively, and mixture transport coefficients are obtained using Wilke’s mixing rule. The species mass diffusion coefficient is determined by assuming a constant Lewis number of 1.4. 

\subsection{Data preprocessing pipeline}

The raw outputs from DSMC, DVM, and continuum CFD solvers differ in spatial format, resolution, variable range, and noise characteristics. A unified preprocessing pipeline is therefore applied to convert these heterogeneous simulation results into standardized learning datasets. This step is essential for fair comparison across neural architectures, since differences in data representation or normalization can strongly affect model performance.

First, all flow fields are mapped onto standardized spatial grids. For the airfoil task, the flow variables are interpolated onto a $128 \times 128$ Cartesian grid. For the cylinder task, an elongated $128 \times 192$ grid is used to capture the extended wake region. For the cavity task, the structured physical domain is retained as $50 \times 50$ while the kinetic moments are organized as multi-channel spatial fields. For the double-cone task, the high-resolution axisymmetric simulation results are mapped onto a sparse and highly anisotropic representation of $17 \times 384$, which is used to evaluate model behavior under shock-dominated and data-limited conditions.

Second, geometry masks are introduced for configurations containing solid bodies. The binary mask M(x) distinguishes fluid regions from the solid or nonphysical regions. During training and evaluation, loss functions and error metrics are computed only over the fluid domain. This avoids forcing models to learn meaningless values inside solid obstacles and prevents invalid interior regions from biasing the comparison among architectures.



Third, variable normalization is applied to reduce scale imbalance across different physical quantities and flow regimes. For variables with moderate dynamic ranges, standard min-max normalization is used. For quantities with large variations, especially in the double-cone task, where pressure and thermodynamic variables may change sharply across shocks, logarithmic dynamic-range compression is applied before normalization. For a positive scalar field $a(\mathbf x)$, the transformed variable can be written as
%
\begin{equation}
    \tilde{a}(\mathbf{x}) = \frac{\mathcal{T}(a(\mathbf{x})) - \min(\mathcal{T}(a))}{\max(\mathcal{T}(a)) - \min(\mathcal{T}(a))} \in [0, 1],
\end{equation}
where
\begin{equation}
    \mathcal{T}(a) = \log_{10}(a + \epsilon),
\end{equation}
and $\epsilon$ is a small positive constant introduced for numerical stability.
The logarithmic compression helps prevent extreme shock-induced jumps from dominating the optimization process.

Finally, all datasets are organized into consistent input-output pairs according to the learning formulation in Section \ref{sec:physical learning}. The input may include geometric descriptors, physical parameters, or operating conditions, while the output consists of task-dependent flow variables. By applying consistent grid mapping, masking, and normalization procedures, TransportBench ensures that model comparisons primarily reflect architectural behavior rather than differences in data preparation.

\section{Benchmarking Protocols and Baseline Models}
\label{sec:baseline}


\subsection{Baseline models}

We evaluate six representative neural architectures that cover different spatial representations and architectural inductive biases commonly used in scientific machine learning for fluid dynamics.

\vspace{0.5em}
\noindent\textbf{Convolutional and latent-representation models}

U-Net is selected as a structured-grid convolutional baseline. Its encoder-decoder architecture, local receptive fields, and skip connections make it effective for capturing multiscale spatial structures and localized gradients. A convolutional autoencoder is also included to evaluate whether non-equilibrium flow fields can be represented through a compact latent bottleneck. This model provides a useful baseline for assessing the compressibility of high-dimensional flow fields and the potential loss of sharp or high-frequency information during latent-space projection.

\vspace{0.5em}
\noindent\textbf{Operator-learning models}

The deep operator network (DeepONet) and Fourier Neural Operator (FNO) are included as representative operator-learning architectures. DeepONet uses a branch-trunk structure to learn nonlinear input-output mappings and supports continuous spatial queries. FNO parameterizes integral kernels in the Fourier domain and captures long-range correlations on structured grids. These two models provide complementary examples of operator learning: DeepONet emphasizes coordinate-based field evaluation, while FNO combines operator learning with a global spectral representation.

\vspace{0.5em}
\noindent\textbf{Attention-based and geometric learning models}

Vision Transformer is included to assess the effectiveness of token-based self-attention for flow-field prediction on structured spatial representations. Point Transformer is selected as a geometric learning baseline that operates on spatial point sets and aggregates information through point-wise attention. These models allow us to evaluate whether global token interactions or flexible point-based representations are beneficial for non-equilibrium flow prediction, especially in settings involving complex geometries, parameter-induced flow variation, or sparse spatial data.

Together, these six architectures span convolutional, latent, operator-learning, spectral, attention-based, and geometric modeling strategies. This selection enables TransportBench to examine how different inductive biases perform under geometry and parameter variations, high-order kinetic-moment prediction, and shock-dominated field reconstruction.


\subsection{Evaluation settings and training protocols}
\label{sec:setting}

To make model comparisons more controlled, TransportBench adopts standardized training and evaluation settings. For Tasks I-III, all baseline models are constrained to approximately one million trainable parameters. This setting is intended to evaluate architectural efficiency under moderate model capacity and to reduce the possibility that performance differences are dominated by model size. For Task IV, which involves sparse data, strong shock discontinuities, and highly anisotropic spatial representation, the parameter budget is increased to approximately 33 million. This setting gives each model sufficient capacity to represent the difficult target fields and helps reveal whether remaining errors are associated with architectural limitations rather than severe under-parameterization.


For Tasks I-III, the datasets are split into training and testing subsets using an 80\%/20\% ratio. The test sets contain unseen samples from the corresponding geometry or parameter distributions. For the airfoil task, this means unseen airfoil geometries generated by CST perturbations. For the cylinder and cavity tasks, the test set contains unseen physical-parameter values or combinations. For the double-cone task, the dataset contains 51 high-fidelity snapshots in total, of which 45 samples are used for training and 6 samples are reserved for testing. The limited number of high-fidelity snapshots makes the task closer to a data-limited shock-dominated prediction setting. All experiments are conducted using the Adam optimizer for 2500 epochs, and the baseline settings are kept consistent across model families unless otherwise specified.


The shock-dominated double-cone task requires additional training stabilization because of the large dynamic range and localized discontinuities in the target fields. For this task, we use a long learning-rate warmup with the OneCycleLR scheduler \cite{smith2018disciplined} to avoid unstable early-stage optimization. A spatial curriculum strategy is also adopted: the model first learns the smoother background field, and the contribution of localized shock regions is increased during later training stages. These strategies are applied consistently to all baseline models in Task IV, so that the comparison remains focused on architectural behavior rather than model-specific tuning.


In addition to the baseline training protocol, Fourier feature injection is evaluated as an ablation in the shock-dominated task. This ablation is designed to test whether explicit high-frequency coordinate embeddings help models resolve sharp shock structures, or whether they introduce oscillatory artifacts, nonphysical noise, or frequency-aliasing effects. The ablation therefore directly connects the model benchmark to the frequency-representation issue discussed in Section \ref{sec:principles}.

\subsection{Masked optimization and evaluation metrics}

For configurations containing solid bodies, the computational domain includes both fluid regions and nonphysical interior regions. To prevent these invalid regions from affecting training and evaluation, we use a binary geometry mask $M(\mathbf x)$, where $M(\mathbf x)=1$ denotes fluid points and $M(\mathbf x)=0$ denotes solid or excluded regions.
During training, models are optimized using a masked mean squared error loss,
%
\begin{equation}
    \mathcal{L}_{\text{MSE}} = \frac{1}{\sum_{i=1}^N \mathbf{M}(\mathbf{x}_i)} \sum_{i=1}^{N} \mathbf{M}(\mathbf{x}_i) \left\| \mathcal{U}_\theta(\mathbf{x}_i) - \mathcal{U}(\mathbf{x}_i) \right\|_2^2,
\end{equation}
where $\mathcal{U}_\theta$ is the model prediction, $\mathcal{U}$ is the reference solution, and $N$ is the number of spatial grid points. 
The MSE loss serves as the optimization objective and penalizes large local deviations more strongly, given its squared-error form. The same mask is used during evaluation so that errors are computed only over the physically valid fluid region.

For quantitative evaluation, we report the masked MSE together with two complementary metrics. The reported MSE is consistent with the training objective and emphasizes larger pointwise deviations because of its squared-error form. The first complementary metric is the masked mean absolute error (MAE),
\begin{equation}
    \text{MAE} = \frac{1}{\sum_{i=1}^N \mathbf{M}(\mathbf{x}_i)} \sum_{i=1}^{N} \mathbf{M}(\mathbf{x}_i) \left\| \mathcal{U}_\theta(\mathbf{x}_i) - \mathcal{U}(\mathbf{x}_i) \right\|_1.
\end{equation}
MAE measures the average pointwise deviation between prediction and reference data.
Compared with the squared training loss, it provides a more direct and interpretable measure of local reconstruction accuracy and is less dominated by a small number of large-error points.

Besides, the masked relative $L_2$ error is defined as
\begin{equation}
    \text{Relative } L_2 = \frac{\sqrt{\sum_{i=1}^{N} \mathbf{M}(\mathbf{x}_i) \left\| \mathcal{U}_\theta(\mathbf{x}_i) - \mathcal{U}(\mathbf{x}_i) \right\|_2^2}}{\sqrt{\sum_{i=1}^{N} \mathbf{M}(\mathbf{x}_i) \left\| \mathcal{U}(\mathbf{x}_i) \right\|_2^2}}.
\end{equation}
The relative $L_2$ error measures the normalized global discrepancy of the predicted field. By scaling the error with the magnitude of the reference solution, it enables more meaningful comparisons across variables and flow regimes with different physical scales. For the double-cone task, the relative $L_2$ error is computed after mapping the normalized predictions and reference fields back to the physical-variable space. This inverse reconstruction includes the inverse logarithmic transformation introduced during preprocessing. The reason is that the logarithmic compression is mainly used to stabilize training and prevent large-magnitude shock-related variables, such as pressure, from dominating the optimization. If the relative $L_2$ error were evaluated only in the log-normalized space, it would measure the discrepancy of the compressed variables and could underestimate physically significant errors near shock-induced peaks. Therefore, evaluating relative $L_2$ in the reconstructed physical space provides a more faithful measure of the relative error associated with strong shocks and large property variations.


In summary, the masked MSE is used as the training objective and is also reported as an evaluation quantity, while MAE and relative $L_2$ provide complementary views of prediction accuracy. MAE characterizes average local accuracy, MSE emphasizes larger pointwise deviations, and relative $L_2$ measures field-level agreement after scaling by the magnitude of the reference solution. For the double-cone task, relative $L_2$ is further evaluated after inverse reconstruction to the physical-variable space so that shock-induced peak errors are not underestimated by the log-normalized representation. 

\section{Numerical Experiments and Analysis}
\label{sec:experiments}

In this section, we evaluate the six baseline architectures introduced in Section \ref{sec:baseline} on the four TransportBench tasks.
The goal is not only to compare prediction errors, but also to examine how different architectural inductive biases behave under distinct non-equilibrium flow conditions. 

\begin{figure}[htbp]
    \centering
    \includegraphics[width=\textwidth]{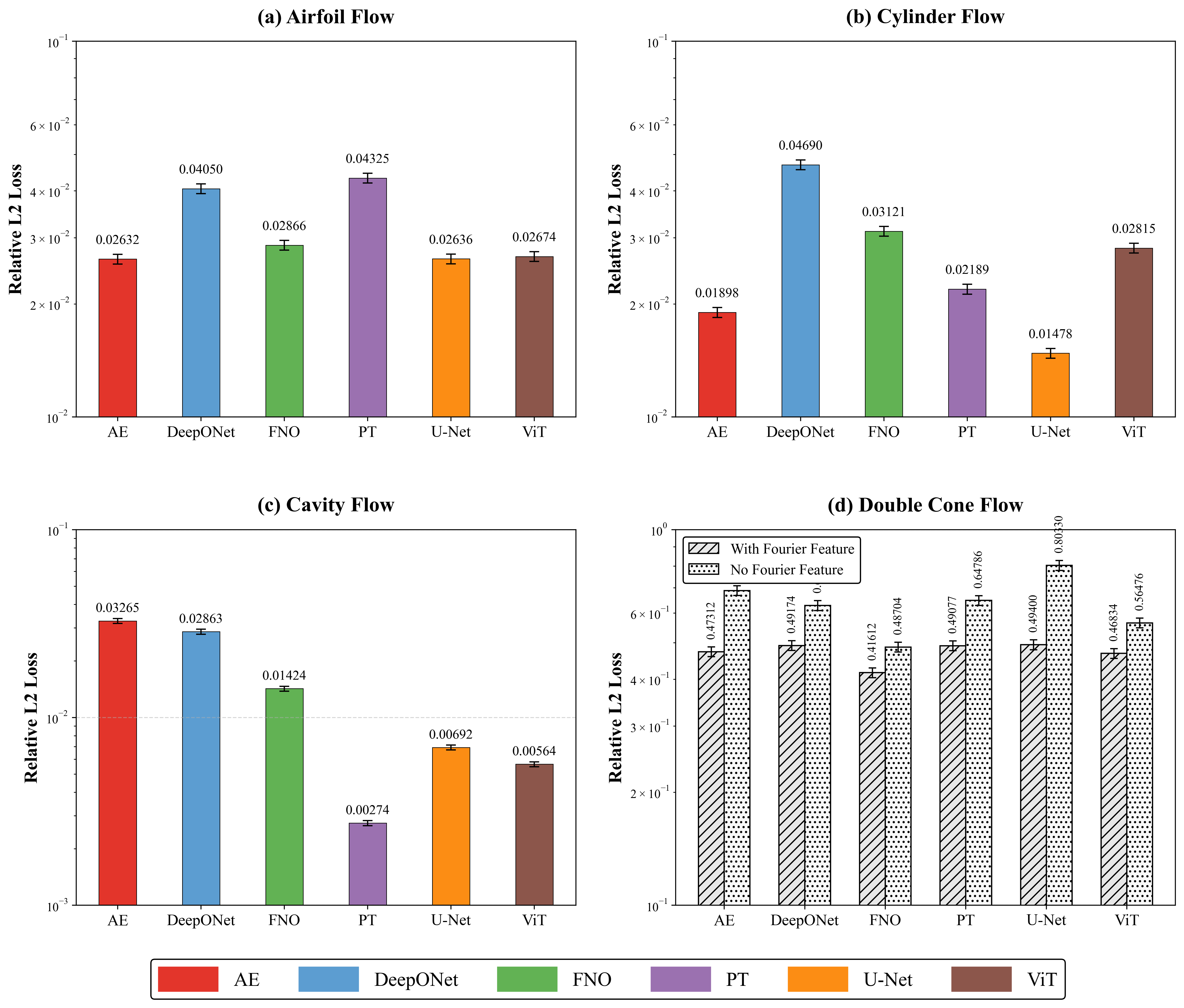} 
    \caption{Quantitative performance overview across the four TransportBench tasks. Subfigures (a)-(c) report the relative $L_2$ error of six baseline architectures on Tasks I-III. 
    The results show regime-dependent model behavior: AE and U-Net achieve low errors in the airfoil task, U-Net performs best in the cylinder task, and PT achieves the lowest error in the cavity task.
    Subfigure (d) reports the Fourier-feature ablation on the double-cone task, showing that high-frequency feature injection reduces relative $L_2$ error for all tested architectures, with architecture-dependent improvement magnitudes.}
    \label{fig:overall_performance}
\end{figure}

Figure \ref{fig:overall_performance} provides an overview of the quantitative results. A clear regime dependence can be observed: no single architecture consistently performs best across all tasks. 
Convolutional and latent-representation models perform favorably in the geometry-varying airfoil task;
U-Net, autoencoder, and Point Transformer show strong performance in the parameter-varying cylinder task;
and Point Transformer achieves the lowest error in the cavity task, followed by Vision Transformer and U-Net.
The double-cone task further highlights the difficulty of resolving localized shock discontinuities under sparse and highly anisotropic representations, and shows that Fourier feature injection has architecture-dependent effects.


\subsection{Task I: airfoil flow with geometry-dependent prediction}

Task I evaluates the ability of models to predict rarefied flow fields around airfoils with geometry variations.
The training and test samples share the same physical operating conditions, while the airfoil shapes vary according to the CST perturbations described in Section \ref{sec:scenarios}.
Therefore, this task focuses on whether a model can learn the relation between solid-boundary geometry and the corresponding non-equilibrium flow structures.

The quantitative results are summarized in Table \ref{tab:task1_airfoil}, which corresponds to Figure \ref{fig:overall_performance}(a). The autoencoder (AE) and U-Net achieve the lowest relative $L_2$ errors among the tested baselines (0.02632 and 0.02636, respectively).
Vision Transformer (ViT) performs closely, with a relative $L_2$ of 0.02674, while FNO remains competitive at 0.02866. 
In contrast, DeepONet and Point Transformer yield larger errors, suggesting that coordinate-based field reconstruction and point-based aggregation can be less effective for this particular geometry-dependent prediction setting.

\begin{table}[htbp]
 \centering
 \caption{Quantitative performance on Task I: airfoil flow with geometry-dependent prediction. MAE, MSE, and relative $L_2$ error are reported for the six baseline architectures on the test set.}
 \begin{tabular}{lccc}
 \toprule
 \textbf{Model} & \textbf{MAE} & \textbf{MSE (L2 Loss)} & \textbf{Relative L2} \\
 \midrule
 AE       & \textbf{0.0120010} & \textbf{0.0002772} & \textbf{0.02632} \\
 DeepONet & 0.0136960          & 0.0006563          & 0.04050 \\
 FNO      & 0.0124002          & 0.0003291          & 0.02866 \\
 PT       & 0.0142392          & 0.0007484          & 0.04325 \\
 U-Net    & 0.0120156          & 0.0002781          & 0.02636 \\
 ViT      & 0.0121577          & 0.0002869          & 0.02674 \\
 \bottomrule
 \end{tabular}
 \label{tab:task1_airfoil}
\end{table}

These results indicate that structured-grid spatial representations are well-suited for the airfoil task. The strong performance of AE and U-Net suggests that local spatial correlations and compact field representations are effective for mapping airfoil-shape variations to the corresponding shock and wake structures. The competitive performance of FNO further indicates that global spectral coupling can also be beneficial, especially for capturing coherent flow-field patterns on structured grids. By contrast, DeepONet lacks explicit local spatial priors, which may limit its ability to resolve geometry-induced shock structures near the leading edge.

\begin{figure}[htbp]
  \centering
  
  \begin{subfigure}{\textwidth}
    \centering
    \includegraphics[width=0.95\textwidth, height=0.42\textheight, keepaspectratio]{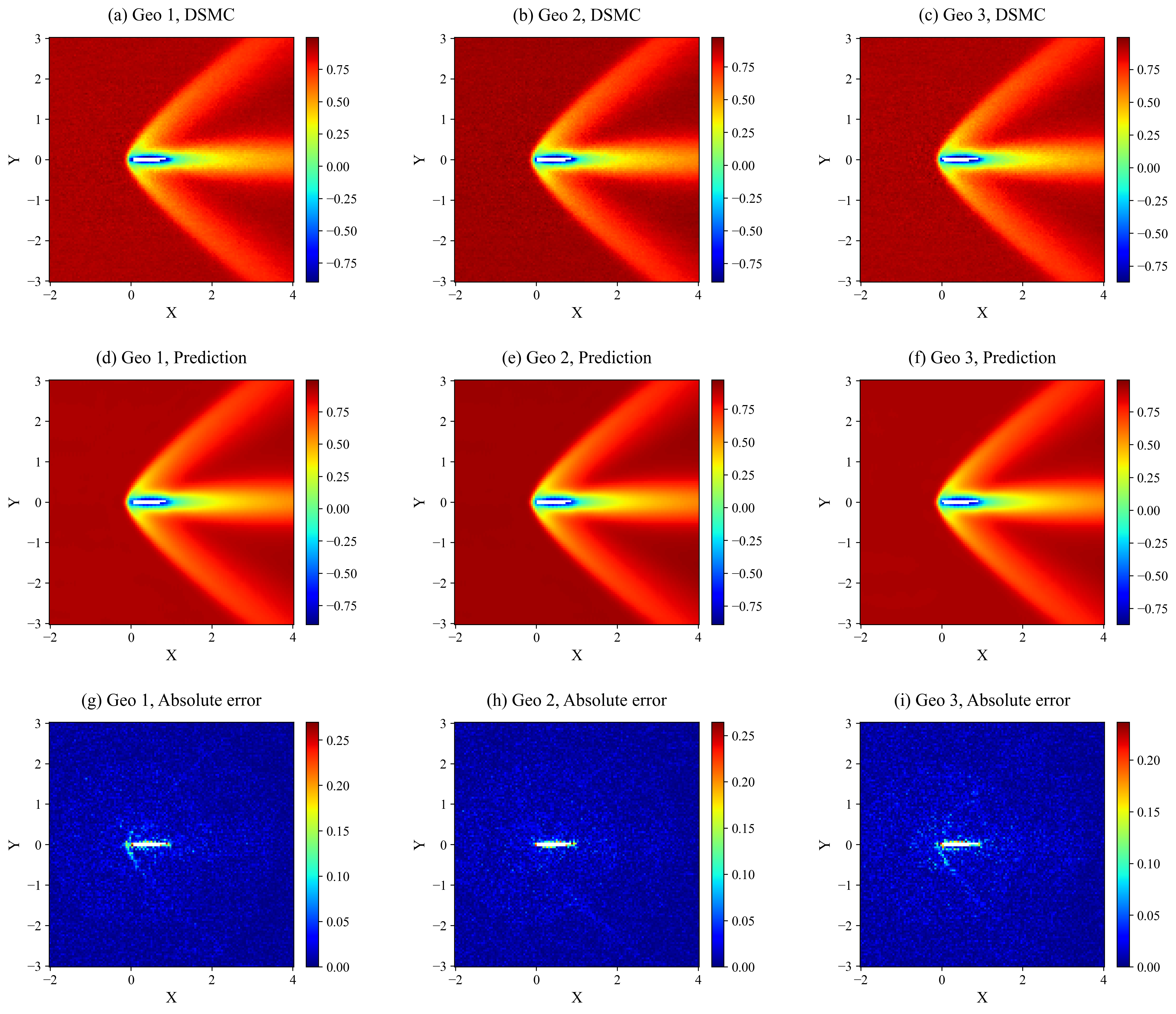 }
    \caption{FNO: lower localized error near shock and wake structures}
  \end{subfigure}
  
  \vspace{1em}
  
  \begin{subfigure}{\textwidth}
    \centering
    \includegraphics[width=0.95\textwidth, height=0.42\textheight, keepaspectratio]{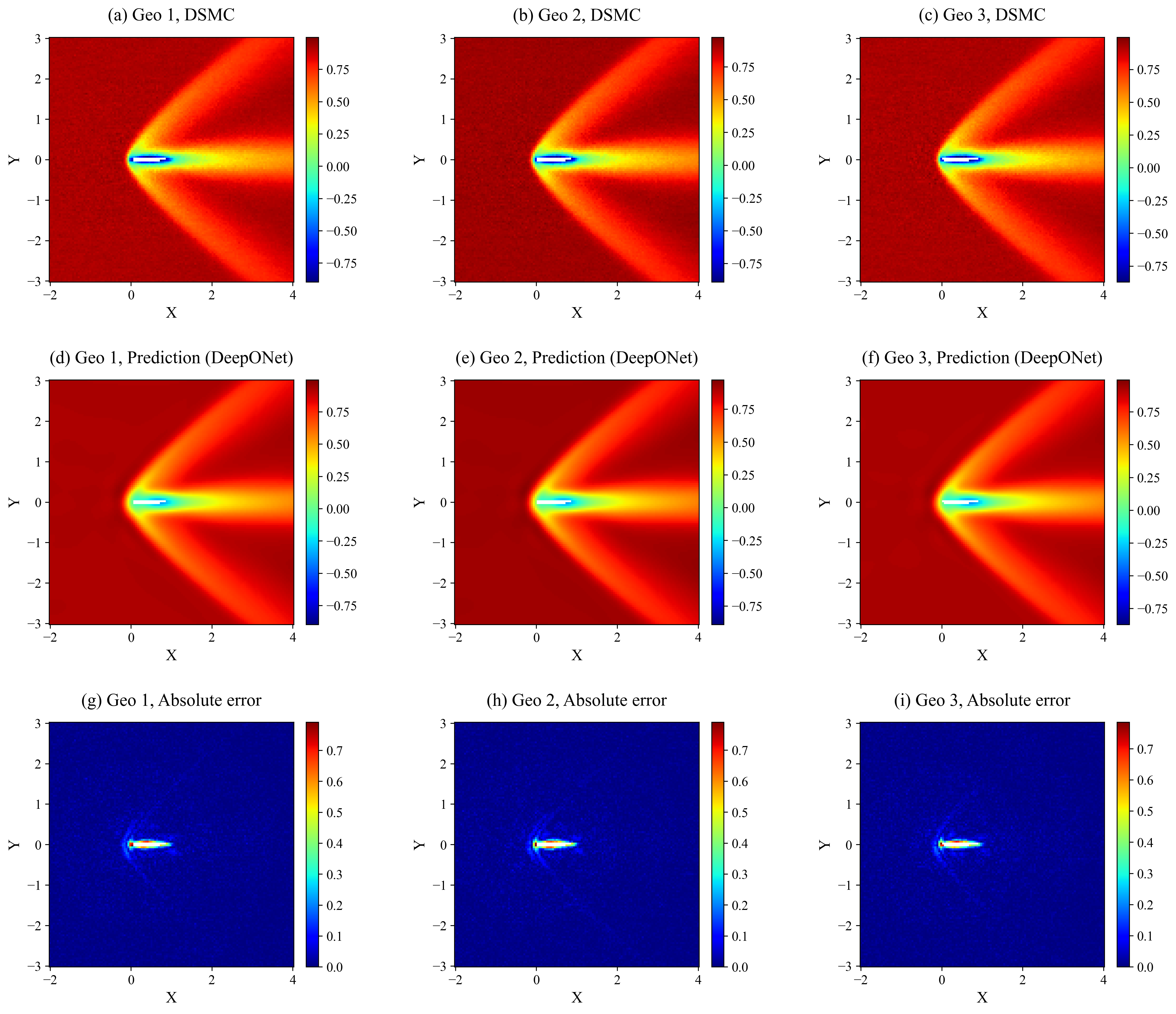}
    \caption{DeepONet: pronounced error concentrated at the leading edge and V-shaped shock front}
  \end{subfigure}
  
  \caption{Comparison on Task I: rarefied airfoil flow. The rows show the DSMC reference solution, model prediction, and absolute error, respectively, for three representative airfoil geometries. FNO reconstructs the geometry-dependent shock and wake structures with lower localized errors, whereas DeepONet exhibits larger discrepancies near the leading edge and along the V-shaped shock front.}
  \label{fig:airfoil_comparison}
\end{figure}

Figure \ref{fig:airfoil_comparison} compares predictions from coordinate-based (DeepONet) and Fourier-based (FNO) representations.
Both models recover the major macroscopic flow features, including the bow shock and downstream wake. Because the reference data are generated by DSMC and therefore contain statistical fluctuations, the smoother model predictions also indicate an implicit denoising effect.
However, clear differences appear near the leading edge and along the shock front. FNO (Figure \ref{fig:airfoil_comparison}a) produces a more accurate reconstruction with relatively uniform error distribution, whereas DeepONet (Figure \ref{fig:airfoil_comparison}b) shows larger discrepancies concentrated around the leading-edge region and the V-shaped shock structure.


Overall, Task I indicates that models with structured-grid spatial priors are well-suited for predicting geometry-dependent flow fields when the main challenge lies in mapping boundary-shape variation to shock and wake structures.

\subsection{Task II: cylinder flow with parameter-dependent prediction}

Task II examines prediction across a broad range of Mach and Knudsen numbers for multi-scale flow around a circular cylinder. Unlike Task I, where the geometry changes but the operating conditions remain fixed, this task requires models to learn how flow topology varies with compressibility and rarefaction. Across the parameter space, the bow shock, wake length, and degree of diffusion change substantially.

The quantitative results are summarized in Table \ref{tab:task2_cylinder}, which corresponds to Figure \ref{fig:overall_performance}(b).
U-Net achieves the lowest errors across all metrics (relative $L_2$ of 0.01478).
AE also performs strongly, reaching a relative $L_2$ of 0.01898, followed by PT at 0.02189.
FNO and ViT show intermediate performance, whereas DeepONet yields the largest error among the tested baselines. These results suggest that parameter-dependent cylinder flow benefits from architectures that can preserve local spatial structures while adapting to large-scale changes in shock and wake topology.


\begin{table}[htbp]
\centering
\caption{Quantitative performance on Task II: cylinder flow with parameter-dependent prediction. MAE, MSE, and relative $L_2$ error are reported for the six baseline architectures across unseen combinations of Mach and Knudsen numbers.}
\begin{tabular}{lccc}
\toprule
\textbf{Model} & \textbf{MAE} & \textbf{MSE (L2 Loss)} & \textbf{Relative L2} \\
\midrule
AE       & 0.0086424          & 0.0002037          & 0.01898 \\
DeepONet & 0.0161770          & 0.0010977          & 0.04690 \\
FNO      & 0.0093879          & 0.0005211          & 0.03121 \\
PT       & 0.0084134          & 0.0002568          & 0.02189 \\
U-Net    & \textbf{0.0065303} & \textbf{0.0001158} & \textbf{0.01478} \\
ViT      & 0.0103630          & 0.0004536          & 0.02815 \\
\bottomrule
\end{tabular}
\label{tab:task2_cylinder}
\end{table}

The strong performance of U-Net indicates that local convolutional priors are effective in this task. Although the wake location and shock structure vary with Ma and Kn, the underlying flow fields remain organized on a structured spatial representation, where local gradients, boundary-layer features, and wake transitions can be captured efficiently by convolutional kernels. The competitive performance of AE further suggests that the dominant flow variations in this dataset are sufficiently coherent to be represented through a compact latent space. Point Transformer also performs well, indicating that point-wise aggregation can adapt to parameter-induced spatial deformation, particularly changes in wake extent and shock position.

\begin{figure}[htbp]
    \centering
    \begin{subfigure}{\textwidth}
        \centering
        \includegraphics[width=0.95\textwidth, height=0.42\textheight, keepaspectratio]{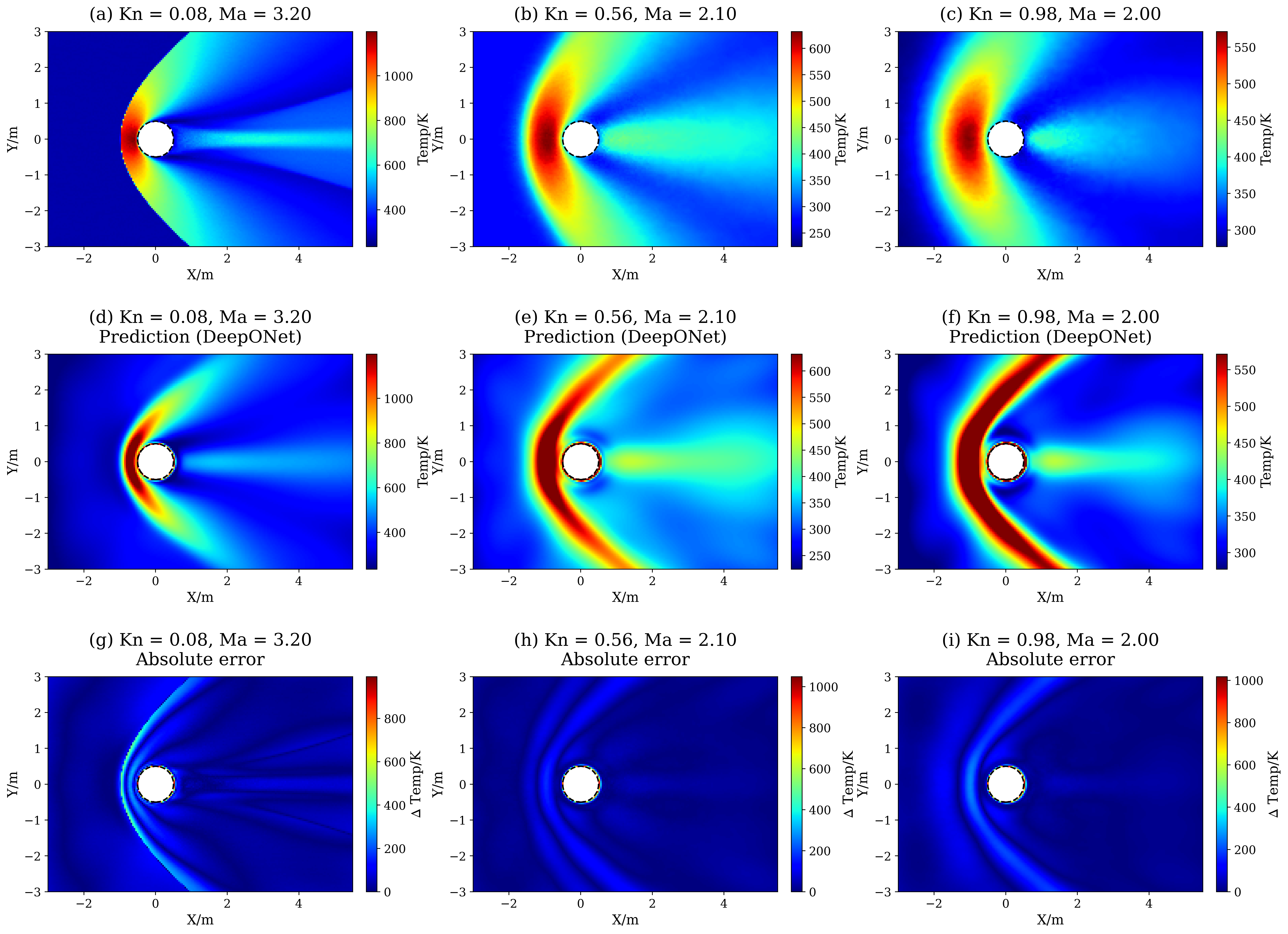}
        \caption{DeepONet: smooth but misaligned shock fronts and wake structures}
    \end{subfigure}
    
    \vspace{1em}
    
    \begin{subfigure}{\textwidth}
        \centering
        \includegraphics[width=0.95\textwidth, height=0.42\textheight, keepaspectratio]{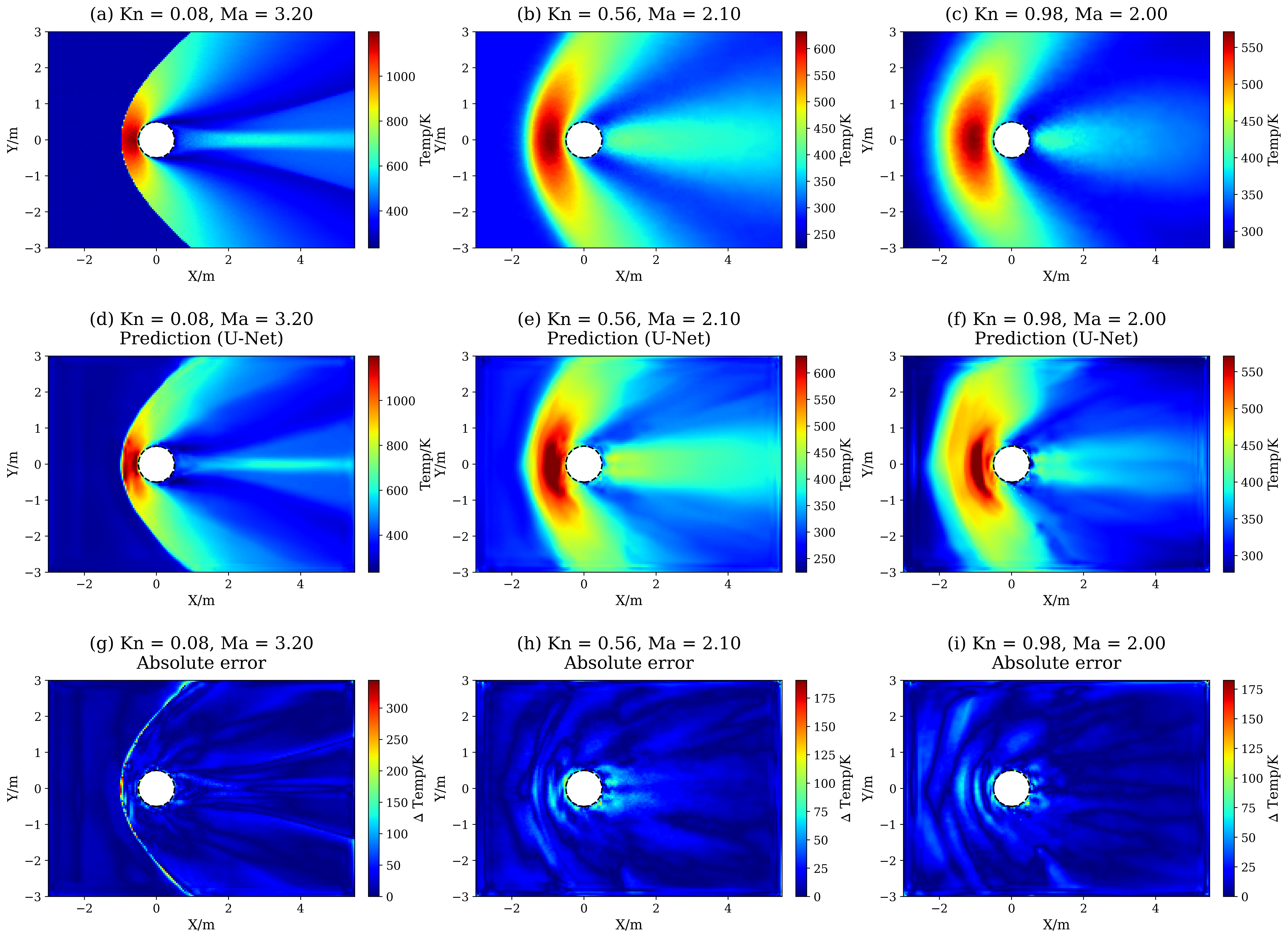}
        \caption{U-Net: localized errors and distorted wake outlines under large parameter variation}
    \end{subfigure}
    
    \caption{Representative predictions on Task II: rarefied cylinder flow. The rows show the DSMC reference solution, model prediction, and absolute error under different combinations of Kn and Ma. DeepONet tends to produce smooth but misaligned shock and wake structures under large parameter variations, while U-Net captures local features but shows errors in the extended wake region. The comparison illustrates different architectural limitations in parameter-dependent flow prediction.}
    \label{fig:cylinder_comparison}
\end{figure}

Figure \ref{fig:cylinder_comparison} compares representative predictions from DeepONet and U-Net under different parameter combinations. DeepONet produces smooth flow fields but tends to misalign shock fronts and wake structures when the operating conditions vary substantially.
This suggests that, in the present architecture, scalar inputs such as Ma and Kn, combined with pointwise coordinate queries, are less effective for reconstructing globally shifting flow topology. U-Net captures the main shock and local flow structures more accurately.
Nevertheless, localized error accumulation can still be observed in the extended wake region under more challenging parameter combinations, indicating that local convolutional priors do not completely remove the difficulty of parameter-dependent topological variation.


These results show that parameter-dependent rarefied-flow prediction requires models to capture both local gradients and global topological changes. Architectures with strong local spatial priors (like U-Net) or flexible spatial point aggregation (like Point Transformer) therefore have clear advantages in this task.

\subsection{Task III: lid-driven cavity flow with high-order kinetic prediction}

Task III evaluates the ability of neural models to predict both the particle distribution functions and the associated macroscopic moment fields.
Compared with Tasks I and II, this task places less emphasis on complex geometry or strong shock structures and instead focuses on whether a model can capture the micro-macro structure of non-equilibrium flow transport.
The distribution function contains the microscopic velocity-space information, while quantities such as density, velocity, stress tensor, and heat flux are obtained from its velocity moments. Therefore, this task provides a direct test of whether neural models can learn not only macroscopic flow patterns, but also the mesoscopic gas-kinetic information.

The quantitative results are summarized in Table \ref{tab:task3_cavity}, which corresponds to Figure \ref{fig:overall_performance}(c).
PT achieves the lowest errors across all metrics with a relative $L_2$ of 0.00274.
ViT and U-Net also perform well, with relative $L_2$ errors of 0.00564 and 0.00692, respectively.
These results indicate that the cavity task does not simply favor a broad model category such as operator learning or grid-based learning. Instead, performance depends on how effectively each architecture represents spatial correlations, boundary-induced non-equilibrium structures, and the coupling between distribution functions and their moment fields.

\begin{table}[htbp]
\centering
\caption{Quantitative performance on Task III: lid-driven cavity flow with high-order kinetic prediction. MAE, MSE, and relative $L_2$ error are reported for prediction of particle distribution functions and associated macroscopic or high-order moment fields.}
\begin{tabular}{lccc}
\toprule
\textbf{Model} & \textbf{MAE} & \textbf{MSE (L2 Loss)} & \textbf{Relative L2} \\
\midrule
AE       & 0.0164398          & 0.0011465          & 0.03265 \\
DeepONet & 0.0087809          & 0.0008484          & 0.02863 \\
FNO      & 0.0026017          & 0.0002289          & 0.01424 \\
PT       & \textbf{0.0011773} & \textbf{0.0000093} & \textbf{0.00274} \\
U-Net    & 0.0042013          & 0.0000511          & 0.00692 \\
ViT      & 0.0028403          & 0.0000370          & 0.00564 \\
\bottomrule
\end{tabular}
\label{tab:task3_cavity}
\end{table}

The strong performance of Point Transformer suggests that point-wise aggregation can be effective for this smooth but physically coupled kinetic problem. Although the cavity geometry is simple, the distribution function and high-order moments are influenced by wall-driven shear, corner effects, and rarefaction. These effects generate spatially distributed correlations that may be captured efficiently by flexible point-based aggregation. The competitive performance of ViT also suggests that token-level global interactions are useful for representing nonlocal correlations in the confined cavity domain. U-Net performs well overall, benefiting from local convolutional priors, but its structured-grid representation might introduce localized artifacts near the corners.

\begin{figure}[htbp]
    \centering
    \begin{subfigure}{\textwidth}
        \centering
        \includegraphics[width=0.95\textwidth]{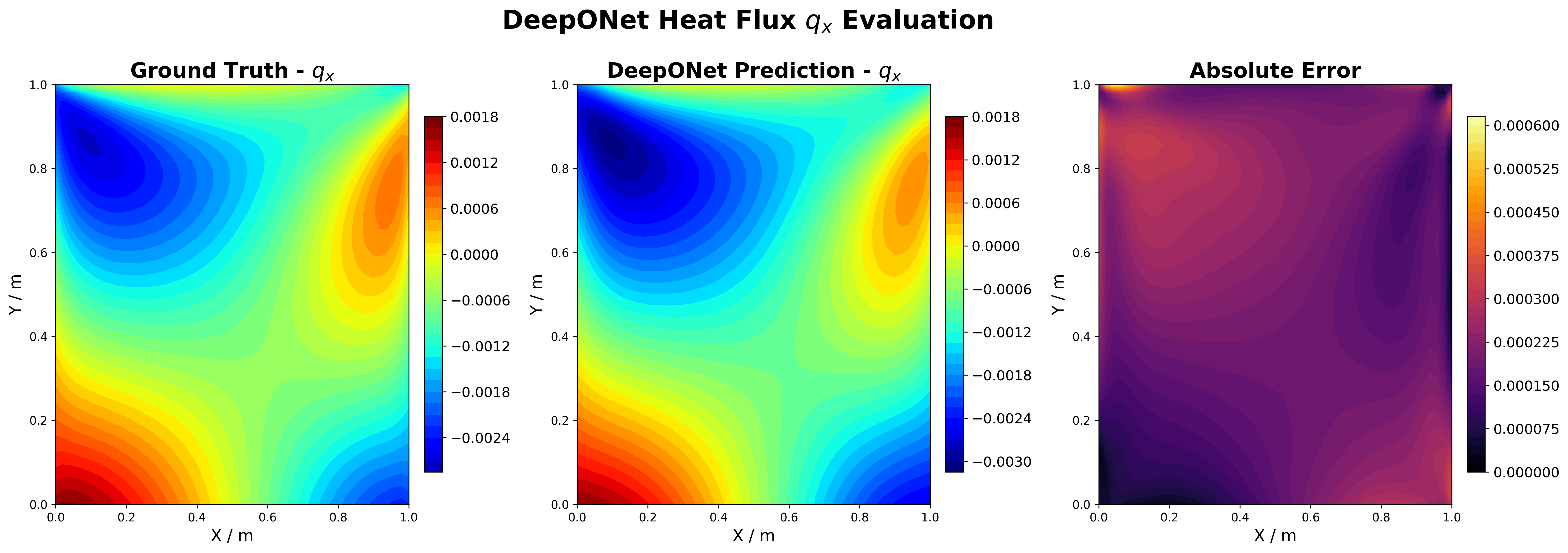} 
        \caption{DeepONet: visually smooth prediction with over-smoothed non-equilibrium variations}
    \end{subfigure}
    \vspace{1em}
    \begin{subfigure}{\textwidth}
        \centering
        \includegraphics[width=0.95\textwidth]{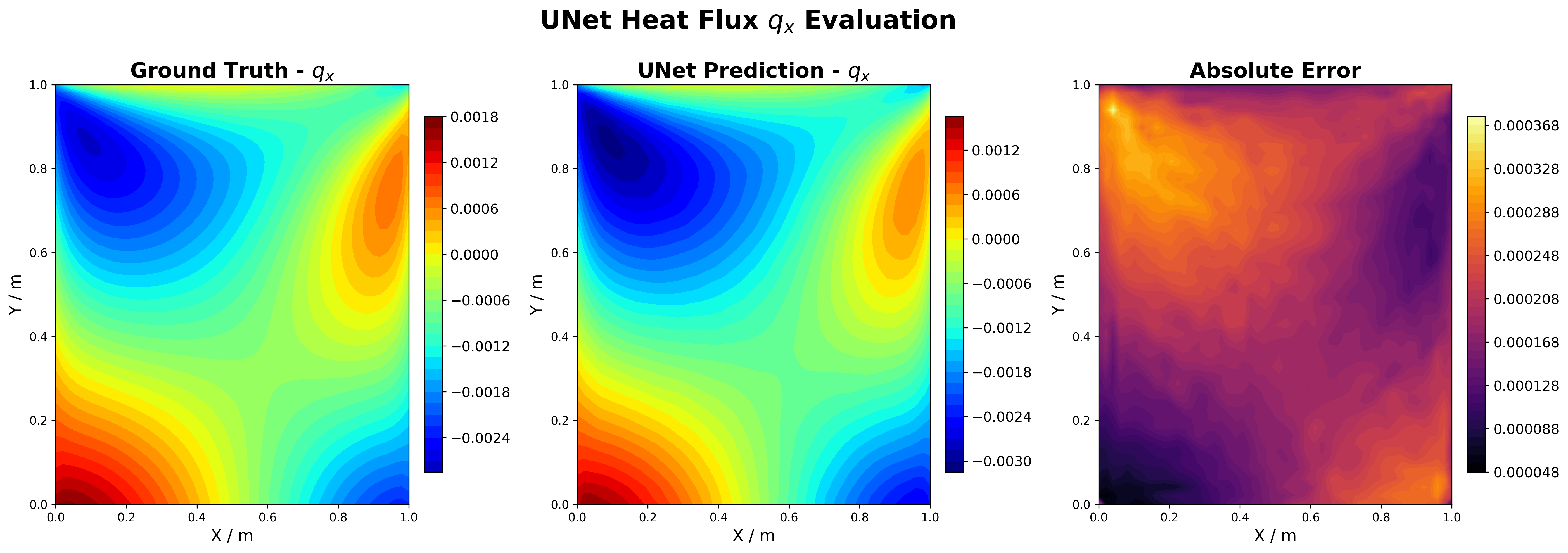} 
        \caption{U-Net: block-like numerical artifacts localized near cavity corners}
    \end{subfigure}
    \caption{Comparison on Task III: lid-driven cavity flow. The figure compares reference, predicted, and absolute-error fields for the heat-flux component $q_x$. DeepONet produces visually smooth predictions but oversmooths some non-equilibrium variations, whereas U-Net achieves much lower overall errors while exhibiting block-like artifacts near the cavity corners. This comparison illustrates two representative error patterns in high-order kinetic-moment prediction.}
    \label{fig:cavity_comparison}
\end{figure}

Figure \ref{fig:cavity_comparison} provides further insight into the representational biases of these models by comparing the heat-flux predictions of DeepONet and U-Net. 
DeepONet produces visually smooth predictions due to its continuous coordinate-based representation, but it over-smooths some non-equilibrium variations and yields relatively large quantitative errors. U-Net achieves a much lower error but exhibits block-like artifacts near the cavity corners. This comparison illustrates that visual smoothness alone is not sufficient for accurate high-order kinetic prediction, while grid-based local reconstruction may introduce discretization-related artifacts in regions affected by boundary singularities.


\begin{figure}[htbp]
  \centering
  \includegraphics[width=0.95\textwidth, height=0.75\textheight, keepaspectratio]{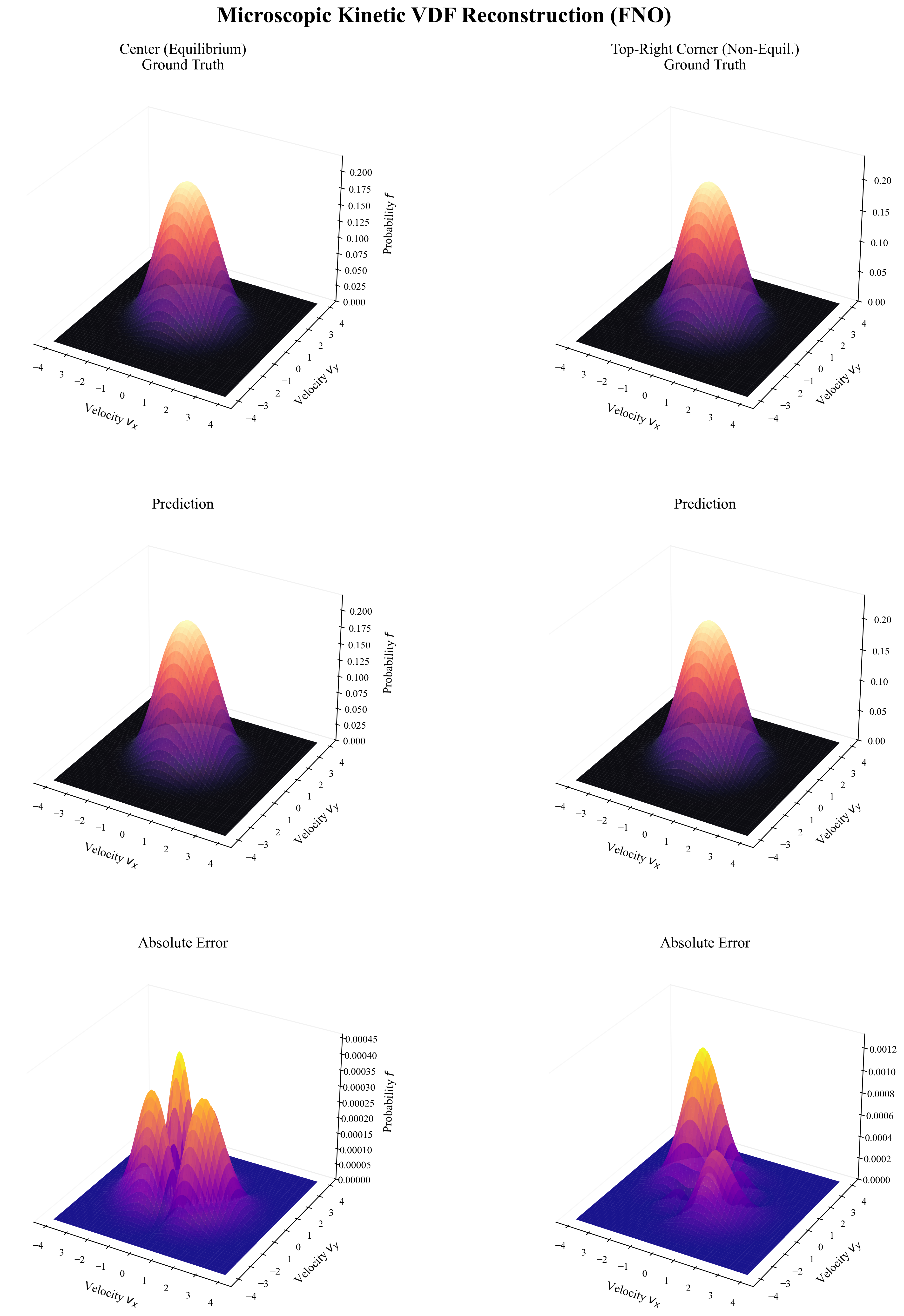}
  \caption{Particle distribution function prediction in the lid-driven cavity flow. The figure compares the reference distribution function, FNO prediction, and absolute error at two representative spatial locations. The left column corresponds to the cavity center, where the distribution remains close to a near-Maxwellian shape. The right column corresponds to the top-right corner, where boundary-driven shear and rarefaction effects lead to an anisotropic and distorted distribution. Since macroscopic quantities and high-order moments are associated with velocity-space integration of the distribution function, this comparison directly evaluates whether the model captures the kinetic structures underlying the predicted moment fields.}
  \label{fig:microscopic_vdf}
\end{figure}

Figure \ref{fig:microscopic_vdf} further examines the prediction results in the velocity space.
It compares the reference and predicted distribution functions at representative spatial locations, i.e., the cavity center and the geometric singularity point at the top-right corner.
At the cavity center (near-equilibrium region), the distribution function remains close to a Maxwellian shape, whereas near boundary-driven shear regions it becomes visibly distorted and anisotropic.
Accurately reproducing these velocity-space structures is important because the high-order moments, such as the stress tensor and heat flux, are direct integrals of the same distribution function. Therefore, agreement at the distribution-function level provides stronger evidence that the model captures the underlying kinetic non-equilibrium, rather than merely fitting smooth macroscopic contours.


Overall, Task III shows that predicting non-equilibrium cavity flow involves a coupled micro-macro learning problem. A successful model should reproduce the distribution function in velocity space and maintain consistency with the macroscopic and higher-order moments derived from it. This makes this task a useful benchmark for evaluating neural solvers on kinetic quantities beyond standard continuum-flow variables.


\subsection{Task IV: double-cone flow with shock-dominated field reconstruction}

Task IV is a demanding task in TransportBench.
It involves high-enthalpy nitrogen flow over a double-cone configuration, with thermochemical non-equilibrium, strong shock discontinuities, shock-boundary-layer interaction, limited training samples, and a sparse anisotropic representation.
The task presents a severe representational challenge for neural models because accurate prediction requires resolving localized shock peaks and steep gradients. 
Besides, the state-to-state model introduces a large number of coupled state-specific variables and stiff source terms associated with vibrational relaxation and dissociation-recombination processes. Therefore, the target fields are compressed manifestations of a high-dimensional thermochemical relaxation process, rather than merely shock contours to be interpolated.


We evaluate the six baseline architectures with and without Fourier feature (FF) high-frequency injection. 
This ablation is designed to examine whether explicit high-frequency coordinate embeddings improve shock resolution and how their effect depends on the underlying model architecture.

\begin{table}[htbp]
\centering
\caption{Quantitative performance on Task IV: double-cone flow. MAE and MSE are computed in the normalized learning space, while relative L2 is computed after inverse reconstruction to the physical-variable space. The ablation study highlights the architecture- and metric-dependent impact of explicit Fourier feature injection in shock-dominated fields.}
\begin{tabular}{llccc}
\toprule
\textbf{Model} & \textbf{Feature} & \textbf{MAE} & \textbf{MSE (L2 Loss)} & \textbf{Relative L2} \\
\midrule
\multirow{2}{*}{AE} & With Fourier & 0.0325 & 0.0058 & 0.47312 \\
 & No Fourier & 0.0313 & 0.0052 & 0.68851 \\
\addlinespace
\multirow{2}{*}{DeepONet} & With Fourier & 0.0397 & 0.0058 & 0.49174 \\
 & No Fourier & 0.0565 & 0.0121 & 0.62819 \\
\addlinespace
\multirow{2}{*}{FNO} & With Fourier & 0.0312 & 0.0060 & \textbf{0.41612} \\
 & No Fourier & 0.0331 & 0.0059 & 0.48704 \\
\addlinespace
\multirow{2}{*}{PT} & With Fourier & 0.0373 & 0.0074 & 0.49077 \\
 & No Fourier & 0.0736 & 0.0210 & 0.64786 \\
\addlinespace
\multirow{2}{*}{U-Net} & With Fourier & 0.0296 & 0.0055 & 0.49400 \\
 & No Fourier & \textbf{0.0259} & \textbf{0.0046} & 0.80330 \\
\addlinespace
\multirow{2}{*}{ViT} & With Fourier & 0.0335 & 0.0057 & 0.46834 \\
 & No Fourier & 0.0378 & 0.0063 & 0.56476 \\
\bottomrule
\end{tabular}
\label{tab:task4_double_cone_ablation}
\end{table}

\subsubsection{Advantage of local convolutional priors}

\begin{figure}[htbp]
    \centering
    \includegraphics[width=0.98\textwidth]{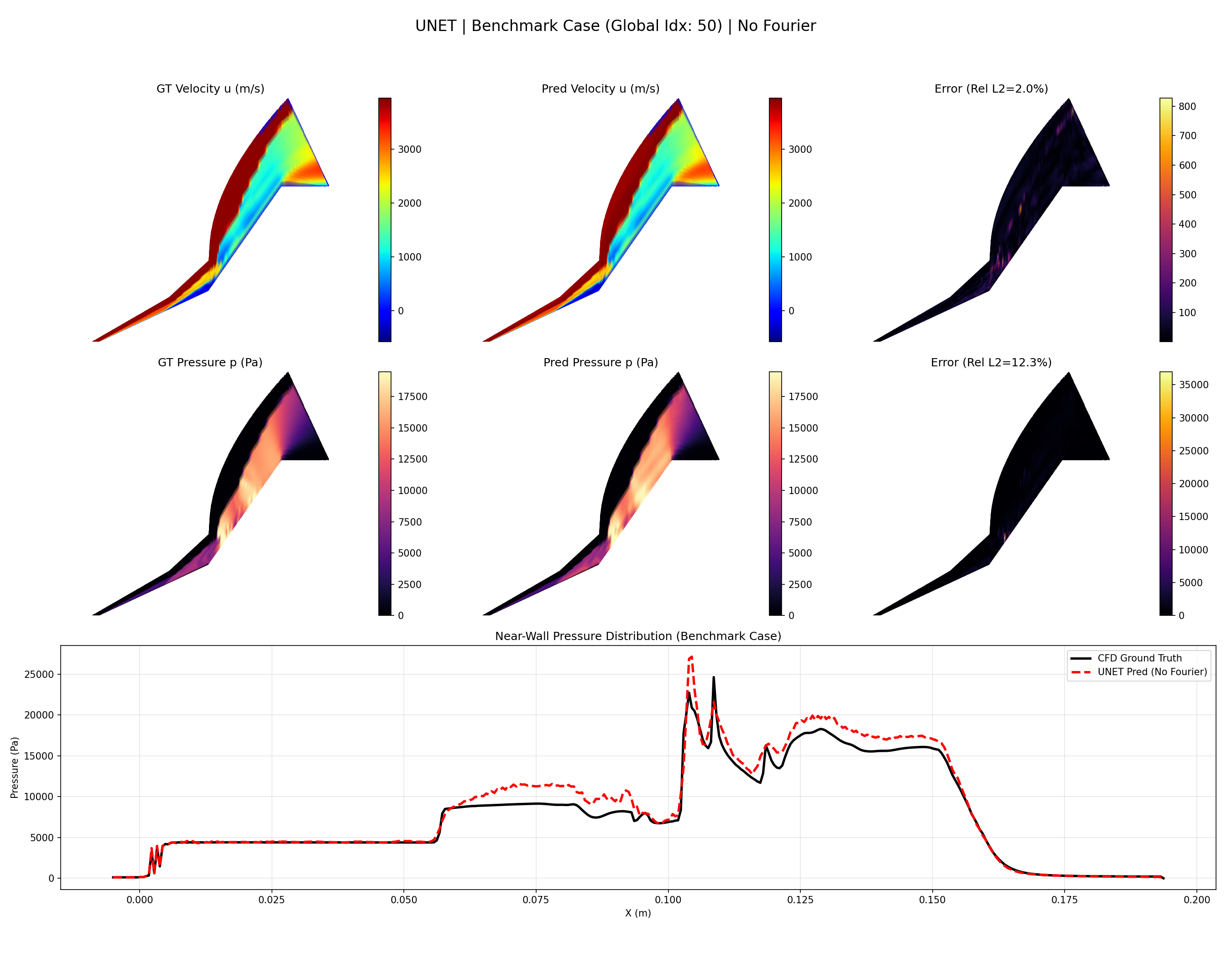} 
    \caption{U-Net prediction on the double-cone task without Fourier feature injection. The figure compares reference and predicted velocity and pressure fields, together with the absolute-error fields and near-wall pressure distribution. U-Net captures the main shock-induced pressure peaks and the overall shock-boundary-layer interaction with relatively low error in this setting, illustrating the advantage of local convolutional priors for shock-dominated field reconstruction.}
    \label{fig:unet_benchmark}
\end{figure}

Among the tested baselines, U-Net without Fourier feature injection achieves the lowest MAE and MSE on the double-cone task, indicating strong local reconstruction capability. However, it does not achieve the lowest relative L2 error. Instead, FNO with Fourier feature injection obtains the best relative L2 performance. 
The near-wall pressure profile in Figure \ref{fig:unet_benchmark} further shows that U-Net without Fourier features captures the main shock-induced pressure peaks and the overall shock-boundary-layer interaction with relatively low error.



This result indicates that local convolutional priors are particularly useful for shock-dominated field reconstruction. The local receptive fields of U-Net help preserve sharp spatial variations, while its encoder-decoder structure and skip connections combine local detail with broader contextual information. 
In the double-cone case, this locality is important not only for resolving shock fronts, but also for representing the rapid downstream variations induced by shock heating and state-specific relaxation.

This finding also highlights an important distinction between smooth-field prediction and shock-dominated prediction. Models that perform well on smooth or globally correlated fields do not necessarily resolve narrow shock structures accurately, especially when training data are scarce and the spatial representation is highly anisotropic.

\subsubsection{Spectral bias and oscillatory artifacts}

\begin{figure}[htbp]
    \centering
    \begin{subfigure}[b]{0.48\textwidth}
        \centering
        \includegraphics[width=\textwidth]{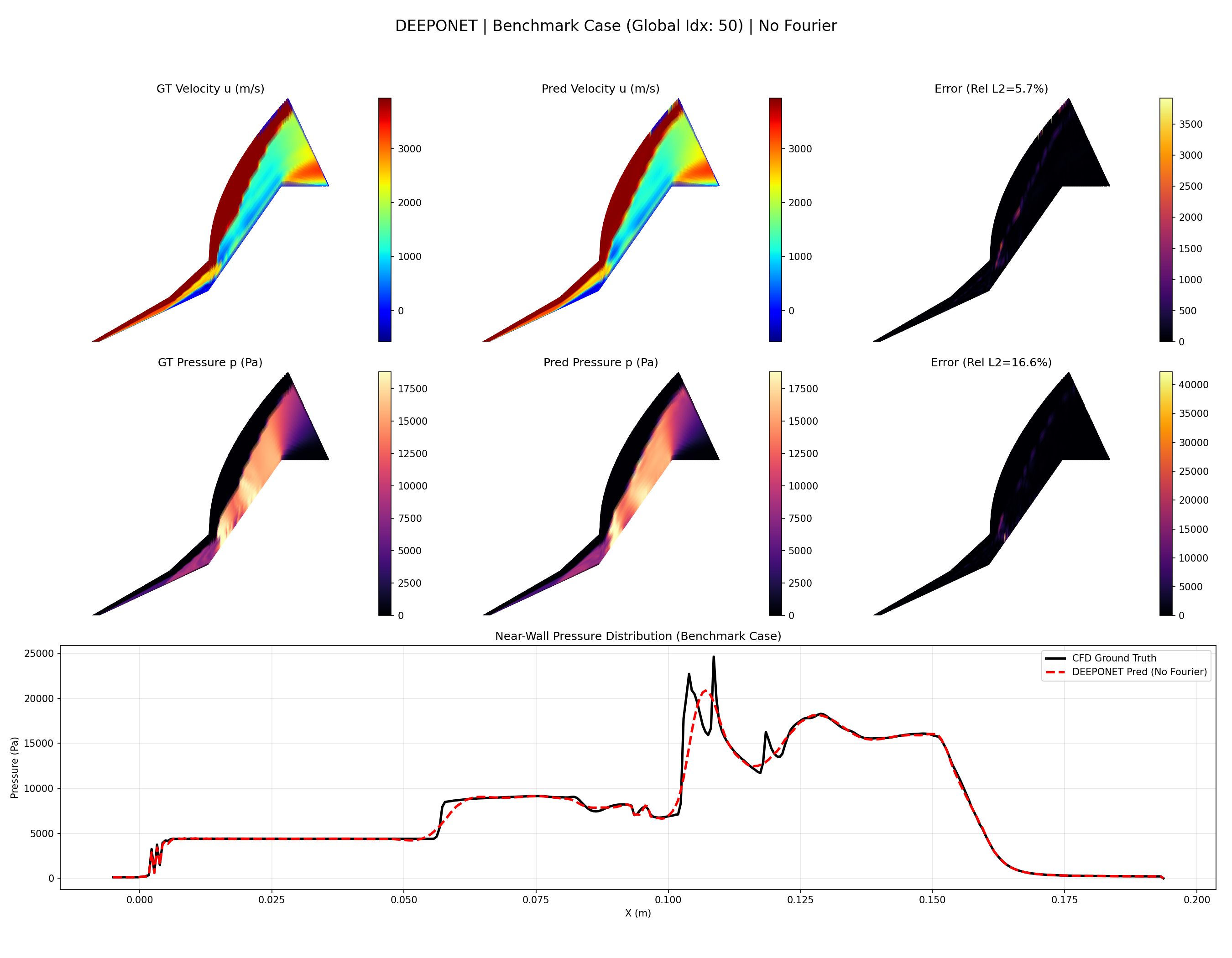}
        \caption{DeepONet (no FF): spectral bias smearing shock peaks}
    \end{subfigure}
    \hfill
    \begin{subfigure}[b]{0.48\textwidth}
        \centering
        \includegraphics[width=\textwidth]{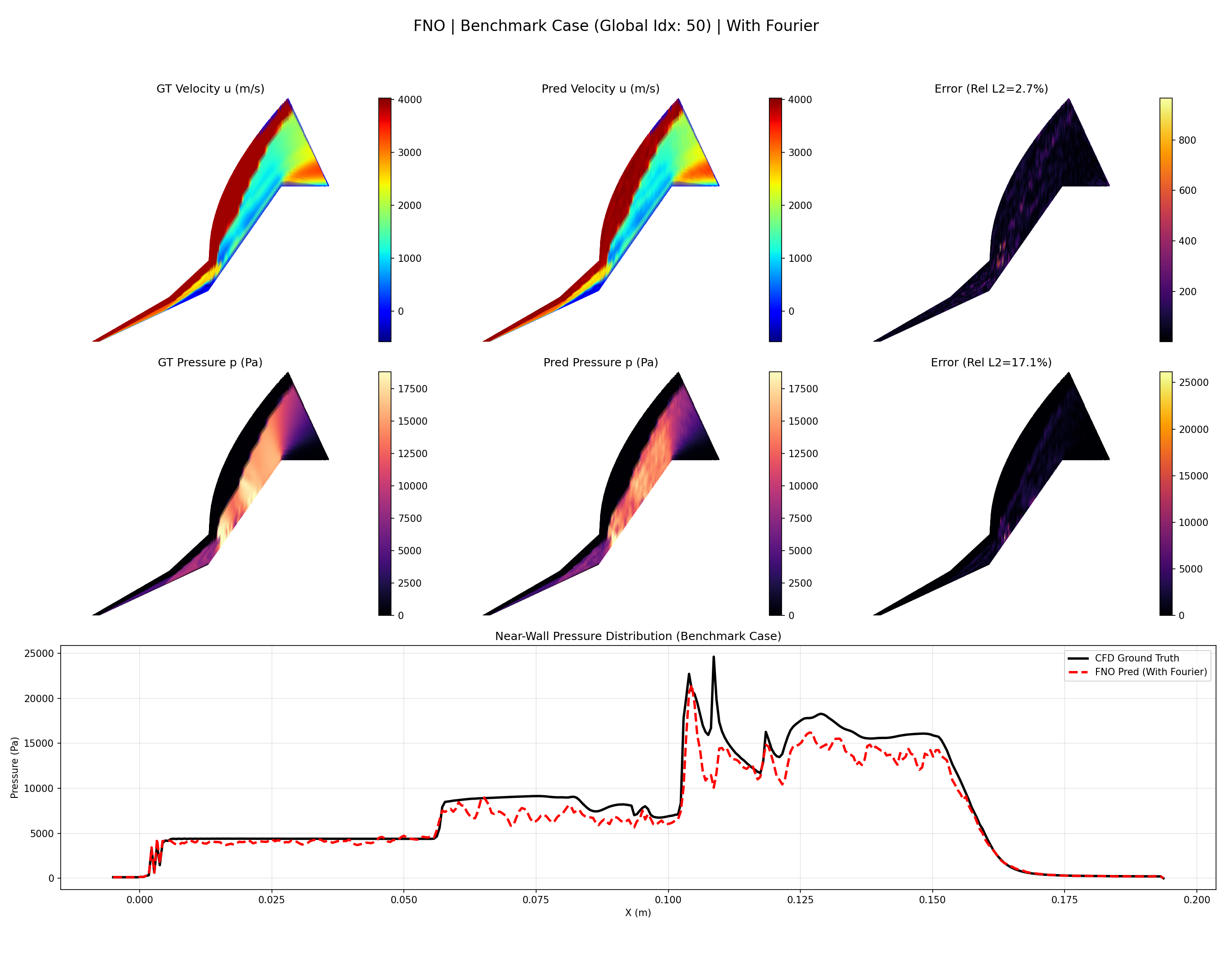}
        \caption{FNO (with FF): Oscillatory artifacts in pressure fields}
    \end{subfigure}
    
    \caption{Representative predictions of DeepONet and FNO on the double-cone task. Panel (a) shows that DeepONet without Fourier feature injection captures the overall flow topology but tends to smooth localized shock-induced pressure peaks. Panel (b) shows that FNO with Fourier feature injection can represent sharper structures in some regions but may introduce oscillatory artifacts in the pressure field. The comparison illustrates two different challenges in shock-dominated prediction: low-frequency smoothing and oscillatory errors near sharp discontinuities.}
    \label{fig:spectral_bias_gibbs}
\end{figure}

Coordinate-based models show limitations in resolving sharp shock peaks. DeepONet without Fourier feature injection captures the overall flow topology but tends to smooth localized pressure maxima near shock-interaction regions, as shown in Figure \ref{fig:spectral_bias_gibbs}(a). This behavior is consistent with the spectral-bias issue discussed in Section \ref{sec:principles}: coordinate-based models may favor smoother, low-frequency mappings and therefore underestimate high-frequency discontinuities.

Fourier-based models face a different challenge. FNO can capture long-range correlations and performs well in several tasks, but its global spectral representation may produce oscillatory artifacts near sharp discontinuities. When Fourier feature injection is applied, FNO improves the reconstruction of some high-frequency structures, but the pressure field may exhibit nonphysical oscillations, as illustrated in Figure \ref{fig:spectral_bias_gibbs}(b). This indicates that a lower global error does not necessarily eliminate local oscillatory artifacts near sharp discontinuities.



These observations suggest that resolving shock-dominated fields requires more than increasing model capacity or adding high-frequency input features. The effectiveness of a model depends on whether its architectural priors are compatible with the localized and discontinuous nature of the target flow field.

\subsubsection{Architecture-dependent effect of Fourier feature injection}

Table \ref{tab:task4_double_cone_ablation} shows that the effect of Fourier feature injection is metric-dependent. In terms of relative L2 error, Fourier features improve all tested architectures. The improvement is most pronounced for U-Net, AE, PT, and DeepONet, indicating that high-frequency coordinate embeddings can enhance physical-space relative field agreement in shock-dominated reconstruction.

However, the same conclusion does not hold uniformly for MAE and MSE. DeepONet, PT, and ViT improve across all three metrics after Fourier feature injection, suggesting that these architectures benefit directly from explicit high-frequency information. FNO obtains the best relative L2 error with Fourier features, although its MSE changes only slightly. In contrast, AE and U-Net exhibit a trade-off: Fourier features reduce relative L2 but increase MAE and MSE. For U-Net in particular, the no-Fourier version remains the best in absolute-error metrics, while the Fourier version substantially improves relative L2.

This metric-dependent discrepancy arises from the different evaluation spaces used by the absolute-error metrics and the relative $L_2$ metric. For stable learning and balanced multi-variable assessment, MAE and MSE are computed in the normalized learning space. This is especially important for the double-cone task, where shock-induced pressure and thermodynamic variables can span several orders of magnitude. Without logarithmic compression and normalization, the training loss and absolute-error metrics would be dominated by a few large-magnitude variables and localized peak regions. Therefore, normalized MAE and MSE mainly reflect the average reconstruction quality over the full masked field, including the broad smooth background.

In contrast, relative $L_2$ is computed after inverse reconstruction to the physical-variable space. For the double-cone task, this inverse reconstruction includes the exponential inverse of the logarithmic transformation. As a result, small errors in the log-normalized representation near shock-induced peaks can correspond to large absolute discrepancies in physical space. Fourier feature injection helps models represent these sharp high-frequency structures more accurately and therefore reduces the physical-space relative $L_2$ error for all tested architectures. However, for architectures with strong local spatial priors, such as AE and U-Net, the same high-frequency embeddings may also introduce small oscillatory errors or background noise in smoother regions, increasing normalized MAE and MSE. Therefore, the results in Table \ref{tab:task4_double_cone_ablation} should be interpreted as a trade-off between full-field normalized reconstruction accuracy and physical-space shock-sensitive relative accuracy.

These results indicate that Fourier feature injection should not be treated as a universally positive or negative modification. Its benefit depends on the interaction between model topology, normalization, and evaluation metric. In shock-dominated thermochemical flows, relative L2, MAE, MSE, and qualitative near-wall behavior should be considered together.

\section{Discussion}


The experiments on TransportBench show that learning non-equilibrium flow transport is strongly shaped by both the underlying physical regime and the model's architectural prior. Beyond reporting prediction errors, the benchmark provides diagnostic evidence on when different model families succeed or fail. We summarize the main implications from the following perspectives.

\vspace{0.5em}
\noindent\textbf{Regime-dependent inductive bias}

A central observation from TransportBench is that model performance is strongly regime-dependent. No single architecture consistently achieves the best performance across all tasks. In the airfoil task, AE and U-Net achieve the lowest errors, while FNO remains competitive, suggesting that structured-grid spatial priors, local convolutional reconstruction, compact latent representation, and global spectral coupling can all be useful for geometry-dependent rarefied-flow prediction. In the cylinder task, U-Net performs best, followed by AE and Point Transformer, indicating that parameter-dependent flow prediction benefits from local spatial priors while also requiring adaptability to large-scale shock and wake deformation. In the cavity task, Point Transformer achieves the lowest error, followed by ViT and U-Net, showing that point-wise aggregation and token-level interactions can be effective for smooth but physically coupled kinetic and moment fields. 
In the double-cone task, U-Net without Fourier feature injection achieves the lowest MAE and MSE, highlighting the value of local convolutional priors for absolute shock-field reconstruction, whereas FNO with Fourier feature injection achieves the lowest relative L2 error. This result shows that shock-dominated reconstruction requires metric-aware assessment rather than a single overall ranking.



These results indicate that model selection should be guided by the dominant structure of the target problem rather than by a presumed universal architecture. Geometry-dependent prediction, parameter-induced topology variation, micro–macro kinetic prediction, and shock-dominated hypersonic reconstruction impose different requirements on neural architectures. TransportBench therefore serves not only as a leaderboard-style benchmark, but also as a diagnostic tool for identifying which inductive biases are appropriate for different non-equilibrium transport regimes.


\vspace{0.5em}
\noindent\textbf{Beyond parameter scaling and black-box benchmarking}



The benchmark also shows that increasing model capacity alone is not sufficient to overcome the difficulty of non-equilibrium flow prediction. This is especially clear in the double-cone task, where the parameter budget is enlarged to reduce severe under-parameterization, yet the observed errors remain strongly architecture-dependent. A model with large capacity can still fail when its spatial representation, frequency bias, or aggregation mechanism is poorly aligned with localized shock structures and thermochemical relaxation effects. Without standardized parameter budgets, grid resolutions, preprocessing strategies, training protocols, and evaluation metrics, performance gains may be difficult to attribute, since they may result from larger models, more aggressive tuning, or higher computational effort rather than from a more suitable architectural inductive bias.

The results further suggest that model assessment should not rely on a single aggregate number. In the double-cone Fourier-feature ablation, for example, high-frequency injection may improve one metric for a specific model while degrading another, indicating that local accuracy, global field agreement, and physical behavior near shocks or boundaries should be considered together. TransportBench addresses these issues by evaluating representative architectures under unified settings, making it easier to separate architecture-related behavior from capacity-driven improvements and to interpret model performance in relation to physical flow structures.


\vspace{0.5em}
\noindent\textbf{High-order variables and frequency-aware representations}

TransportBench extends conventional fluid benchmarks by including kinetic-level quantities and high-order non-equilibrium statistics. In the cavity task, models are evaluated on both particle distribution functions and the macroscopic or high-order moment fields derived from them. These quantities are directly connected to the velocity-space structure of the Boltzmann solution and are more sensitive to rarefaction and non-equilibrium effects than standard macroscopic variables. Accurate prediction therefore requires more than reproducing density, velocity, or temperature fields; it requires preserving the micro–macro consistency between the distribution function and its moments.

At the same time, the double-cone task shows that shock-dominated hypersonic flows pose a different but equally important representational challenge. Models that favor smooth or low-frequency mappings may capture the overall flow topology but under-resolve localized shock peaks. Explicit Fourier feature injection can mitigate this tendency in some cases, such as reducing the relative $L_2$ error of DeepONet, but it is not universally beneficial. 
In the double-cone results, explicit Fourier feature injection reduces relative $L_2$ errors for all tested architectures and improves all three metrics for DeepONet, Point Transformer, and ViT. However, it also produces metric-dependent trade-offs for AE and U-Net, where relative L2 improves while MAE and MSE deteriorate. These findings indicate that future models for non-equilibrium flow transport should be both kinetic-aware and frequency-aware, while also being evaluated with multiple complementary metrics.

\section{Conclusion}


In this work, we introduced TransportBench, a benchmark framework for non-equilibrium flow transport that integrates high-fidelity datasets with standardized model evaluation protocols. The benchmark is designed to address two related needs in scientific machine learning for fluid dynamics: the lack of representative datasets for non-equilibrium gas transport and the need for controlled comparison of neural architectures across different physical regimes.

TransportBench covers a broad range of transport phenomena, including rarefied and continuum flows, low-speed and hypersonic conditions, inert and reacting gases, translational and internal-energy non-equilibrium, and both smooth and shock-dominated field structures. The four benchmark tasks are designed to assess geometry-dependent prediction, parameter-dependent prediction, distribution-function and high-order kinetic prediction, and shock-dominated hypersonic reconstruction. By including non-equilibrium quantities such as distribution functions, stress, and heat flux, the current work extends existing fluid benchmarks beyond standard macroscopic variables and provides a more stringent test for learning non-equilibrium transport.

Through systematic evaluation of representative convolutional, latent-representation, operator-learning, attention-based, and geometric models, we showed that model performance is strongly regime-dependent. No single architecture performs uniformly best across all tasks. Instead, different inductive biases become advantageous under different physical and representational conditions. 
The results also show that explicit high-frequency feature injection has a non-uniform effect in shock-dominated flows. In the double-cone benchmark, Fourier features reduce relative L2 errors across the tested architectures, but their influence on MAE and MSE remains architecture-dependent. Thus, high-frequency augmentation should be evaluated jointly with global normalized errors, absolute pointwise errors, and qualitative physical behavior, since improved aggregate accuracy may still coexist with local oscillatory artifacts or nonphysical noise.

Overall, TransportBench provides not only a reusable dataset for non-equilibrium gas dynamics, but also a diagnostic benchmark for understanding how scientific machine learning models behave across multiple levels of gas-transport physics. We expect it to support the development of more robust, physics-aware, and regime-adaptive neural solvers for rarefied flows, high-enthalpy hypersonic flows, and broader non-equilibrium transport problems.

\section*{Acknowledgements}

The current research is funded by Strategic Priority Research Program of Chinese Academy of Sciences (XDB0620403), Chinese Academy of Sciences Project for Young Scientists in Basic Research (YSBR107), National Science Foundation of China (12302381, 12572340), and Beijing Natural Science Foundation (L252039).

\bibliographystyle{elsarticle-num} 
\bibliography{refs}                

\newpage
\appendix
\section{Numerical Implementation and Hyperparameters}
\label{app:hyperparameters}

To ensure complete reproducibility of the TransportBench evaluation, we detail the exact hyperparameter configurations and architectural specifications for all baseline models evaluated in this study. 

To decouple intrinsic architectural advantages from brute-force parameter scaling, all models in Tasks I-III are bottlenecked to approximately $1 \times 10^6$ (1M) parameters. For Task IV, the budget is scaled to approximately $33 \times 10^6$ (33M) parameters to combat severe physical discontinuities. The detailed architectural hyperparameters (e.g., hidden dimensions, number of layers/modes) configured to meet these budgets, alongside their specific optimization settings, are summarized in Table \ref{tab:hyperparameters}. 

All models are trained for 2500 epochs using the Adam optimizer. For Task IV, the \texttt{OneCycleLR} scheduler dictates the learning rate warm-up and annealing phases.

\begin{table}[htbp]
    \centering
    \caption{Hyperparameter configurations of the six baseline architectures used in TransportBench. The table lists the parameter budget, learning rate, batch size, and key architectural settings for Tasks I-III. For Task IV, the parameter budget is increased as described in Section \ref{sec:setting}, while the same architecture families and training framework are retained.}
    \renewcommand{\arraystretch}{1.3}
    \resizebox{\textwidth}{!}{
    \begin{tabular}{@{} l c c c l @{}}
        \toprule
        \textbf{Model} & \textbf{Params (Standard)} & \textbf{Learning Rate} & \textbf{Batch Size} & \textbf{Key Architectural Configurations} \\
        \midrule
        \textbf{DeepONet}      & $\sim$1.05M & $1 \times 10^{-3}$ & 32 & Branch layers: 4, Trunk layers: 4, Hidden dim: 128 \\
        \textbf{FNO}           & $\sim$0.98M & $1 \times 10^{-3}$ & 32 & Fourier modes: 12, Width: 32, FFT layers: 4 \\
        \textbf{U-Net}         & $\sim$1.10M & $5 \times 10^{-4}$ & 16 & Encoder/Decoder blocks: 4, Base channels: 32 \\
        \textbf{ViT}           & $\sim$1.02M & $1 \times 10^{-4}$ & 16 & Patch size: $8\times 8$, Transformer layers: 6, Heads: 8 \\
        \textbf{AutoEncoder}   & $\sim$1.00M & $1 \times 10^{-3}$ & 32 & Conv layers: 5, Latent dim: 128, Activation: GELU \\
        \textbf{Point Trans.}  & $\sim$1.15M & $5 \times 10^{-4}$ & 8  & $k$-nearest neighbors: 16, Attention layers: 4 \\
        \bottomrule
    \end{tabular}
    }
    \label{tab:hyperparameters}
\end{table}

\end{document}